\documentclass[%
 reprint,
superscriptaddress,
 amsmath,amssymb,
 aps,
prb,
]{revtex4-2}
\usepackage{mathrsfs}
\usepackage{orcidlink}
\hypersetup{
     colorlinks=true,  
     linkcolor=blue,   
     citecolor=blue,
     urlcolor=blue
}
\usepackage{graphicx}
\usepackage{dcolumn}
\usepackage{bm}

\begin{document}

\preprint{APS/123-QED}

\title{Probing kinetic enhancement of fusion reactivity in turbulent hot spots}

\author{Yao Guo\orcidlink{0009-0009-7523-5887}}
\affiliation{%
State Key Laboratory of Dark Matter Physics, Key Laboratory for Laser Plasmas, Department of Physics and Astronomy, Shanghai Jiao Tong University, Shanghai 200240, People’s Republic of China
}%

\author{Dong Wu\orcidlink{0000-0001-5738-5739}}%
 \email{dwu.phys@sjtu.edu.cn}

\affiliation{%
State Key Laboratory of Dark Matter Physics, Key Laboratory for Laser Plasmas, Department of Physics and Astronomy, Shanghai Jiao Tong University, Shanghai 200240, People’s Republic of China
}%
\affiliation{Collaborative Innovation Center of IFSA (CICIFSA), Shanghai Jiao Tong University, Shanghai 200240, People’s Republic of China}

 \author{Jie Zhang\orcidlink{0000-0001-7821-4808}}
 \email{jzhang1@sjtu.edu.cn}
 
\affiliation{%
State Key Laboratory of Dark Matter Physics, Key Laboratory for Laser Plasmas, Department of Physics and Astronomy, Shanghai Jiao Tong University, Shanghai 200240, People’s Republic of China
}%

\affiliation{Collaborative Innovation Center of IFSA (CICIFSA), Shanghai Jiao Tong University, Shanghai 200240, People’s Republic of China}

\affiliation{Institute of Physics, Chinese Academy of Sciences, Beijing 100190, People’s Republic of China}

\date{\today}

\begin{abstract}
Traditionally, fusion reactivity in thermonuclear plasmas has been calculated by assuming a local Maxwellian ion distribution. However, recent theoretical work [Phys. Rev. Lett. 135, 155101 (2025)] suggests that turbulence in plasmas can generate non-Maxwellian tail distributions, thereby enhancing reactivity. In this paper, we investigate this effect through numerical simulations of a sinusoidal shear flow. By comparing steady-state distributions obtained with the Bhatnagar–Gross–Krook (BGK) and Fokker-Planck (FP) collision operators, respectively, we demonstrate that the BGK model overestimates the reactivity enhancement while the FP operator gives a much more modest enhancement that is nearly halved under typical ICF parameters. Particle-in-cell (PIC) simulations incorporating nuclear reactions are also conducted, which reveal that the combined effects of preferential ion heating during shear flow dissipation and tail enhancement can even amplify the reactivity enhancement to be larger than the steady-state prediction.

\end{abstract}

\maketitle


\section{\label{sec:level1}Introduction}

Inertial confinement fusion (ICF) is a scheme in which fuel targets are irradiated, heated and compressed by intense lasers to form a dense hot spot and achieve fusion ignition \cite{nuckolls1972laser,Lindl2004,betti2016inertial}. Traditionally, the evolution of the fuel target has been primarily modeled using single-fluid  radiation-hydrodynamic (rad-HD) approaches \cite{Marinak2001,Clark2019}. This framework rests on the assumption that the fuel plasma achieves thermal equilibrium (i.e. relaxes to a Maxwellian distribution) on a timescale much shorter than any other relevant dynamical timescales \cite{spitzer2006physics}. Under this assumption, fusion reactivity within the hot spot is a local function dependent solely on reactant densities and temperatures; thus, maximizing the conversion of laser energy into thermal energy is, in principle, optimal for fusion.

Consequently, any residual kinetic energy (in this paper, kinetic energy refers to the macroscopic kinetic energy of the fluid) in the hot spot is considered wasted and non-productive for fusion reactivity \cite{Kritcher2014}. Furthermore, remnant kinetic energy often results from hydrodynamic instabilities in the compression process, such as Richtmyer-Meshkov (RM) and  Rayleigh-Taylor (RT) instabilities \cite{ZHOU20170,ZHOU20171,Guo2024}. These instabilities can reduce the hot spot symmetry and lead to deformation, breakup, and radiative cooling \cite{Springer_2019}, which are destructive to hot spot formation and confinement. Accordingly, for years, the National Ignition Facility (NIF) campaigns \cite{Kline_2019}, as well as various alternative ignition schemes (e.g., fast ignition \cite{Tabak2005} and double-cone ignition \cite{Zhang2020}), have striven to suppress hydrodynamic instabilities and the resultant turbulence. Substantial progress has been made in this direction, with fusion gains exceeding unity having been achieved several times at NIF \cite{zylstra2022burning,Abu2024Achievement}.

Despite these successes, single-fluid simulations frequently fall short of accurately predicting experimental observations. These discrepancies are often attributed to kinetic and multi-species effects \cite{Rinderknecht2014,rinderknecht2018kinetic,Yin2019,hartouni2023evidence,Higginson2025Evidence} that lie beyond the single-fluid approximation. Specifically, single-fluid models consistently fail to explain the apparent ion temperature anomalies (e.g., inferred $T_{DD}<T_{DT}$ \cite{Murphy2014,Mannion2023}) and anomalous neutron yield \cite{Taitano2018} observed in experiments. This arises partly because HD models calculate reactivity assuming a local Maxwellian distribution based solely on local temperature and density. In contrast, kinetic theory determines fusion reactivity by integrating over the full velocity distribution, where the most significant contribution comes from the high-energy tail. Because these supra-thermal ions are weakly collisional, they are highly susceptible to non-Maxwellian distortions. Consequently, incorporating such kinetic effects can lead to yield predictions that deviate significantly from standard HD models. Notable efforts in this direction include studies of Knudsen-layer loss \cite{Petschek_1979,Molvig2012Knudsen,Albright2013,Schmit2013} and $\alpha$-particle knock-on effects \cite{Fisher_1994,Ballabio1997,XUE2025359}.

Recently, Fetsch and Fisch (FF hereafter) \cite{fetsch2025enhancement,fetsch2025analytical} proposed that turbulence and shear flows within ICF hot spots may counter-intuitively favor fusion ignition through the spatial transport of tail ions. This mechanism is actually analogous to collisionless shear acceleration, which was proposed decades ago in the astrophysical context \cite{Rieger_2004}. In collisionless shear flows, particles randomly traverse the shear and scatter off magnetic islands, thereby gaining a velocity difference and being accelerated as the process repeats. In a strongly collisional environment like ICF hot spots, this effect is inhibited for most particles  since their mean free path $\lambda$ is much shorter compared to the velocity gradient length $L$, i.e. the Knudsen number $\text{Kn}\equiv   \lambda/L\ll1$ is small. However, since $\lambda\propto v^4$, a few fast tail particles can break this constraint and surf across the shear. Consequently, these tail particles gain a relative velocity increase and march further into the high-velocity region of the phase space (more precisely, approximately half of the tail particles are accelerated across the shear, while the remaining half are decelerated).

Building on this, FF developed an analytical theory to describe the shear flow reactivity enhancement (SFRE) in subsonic turbulent flows. By employing a perturbation expansion in Mach number on a modified Bhatnagar–Gross–Krook (BGK) equation, they derived explicit expressions for the perturbed ion distribution up to the second order. These perturbed distributions are then used to derive an explicit (though complicated) formula for the reactivity enhancement factor $\Phi$. Their calculations demonstrate that under typical ICF parameters, the reactivity can be enhanced several-fold compared to that of the local Maxwellian distribution. Crucially, the SFRE reactivity can exceed the values obtained under the assumption of complete thermalization of turbulent kinetic energy (TKE). This in principle enables novel ICF designs with lower ignition temperatures, smaller hot spot radii, and in consequence, smaller driving energy, as long as turbulence in the hot spot can be driven at a lower expense.

Despite these profound insights, the impact of SFRE on the dynamic evolution of a realistic hot spot remains an open question. Firstly, the analysis in FF relies on a modified BGK operator, which oversimplifies the phase-space scattering process. Energy diffusion and pitch-angle scattering in reality can suppress the formation of anisotropic high-energy tails, hindering SFRE. Secondly, FF assumes a static background flow, relying on the assumption of an infinite timescale separation between the fast tail ions and the bulk flow evolution. In a real hot spot, however, the viscous dissipation of turbulence, the thermalization of the plasma, and the self-heating from $\alpha$-particles occur on finite and comparable timescales. This necessitates a fully time-dependent, self-consistent evaluation of the flow decay and feedback loops.

These unresolved issues motivated this work. In this paper, we investigate the SFRE effect in a systematic way. The paper is organized as follows. In Sec.\ \ref{sec:theo}, following FF, we provide a brief review of the theoretical basis underlying this problem. In Sec.\ \ref{sec:tail}, we investigate the steady-state tail distribution with both the BGK operator and the Fokker-Planck (FP) operator, where it is shown that the BGK operator significantly overestimates SFRE. With the FP operator, we propose SFRE optimal in a smaller parameter space. In Sec.\ \ref{sec:pic}, we conduct particle-in-cell (PIC) simulations of the turbulent hot spot, incorporating fusion reactions. Direct evidence of an enhanced ion tail distribution and increased reactivity in turbulent hot spots is presented. In Sec.\ \ref{sec:con}, we draw a conclusion and discuss some outlooks.

\section{Theoretical basis}\label{sec:theo}
In this section, we briefly review the theoretical foundation underlying the SFRE effect.
\subsection{Fusion reactivity}
In a thermonuclear plasma, the fusion reactivity  between two reactant ion species $i$ and $j$, $\Sigma[f_i,f_j]\equiv\langle \sigma v\rangle_{ij}$, is given by
\begin{equation}
\langle\sigma v\rangle_{ij}=\frac{2-\delta_{ij}}{2}\iint d^3\bm vd^3\bm v'\sigma(v_r)v_rf_i(\bm v)f_j(\bm v'),\label{eq:1}
\end{equation}
where $v_r=|\bm v-\bm v'|$ is the magnitude of the relative velocity between the two reactants. The fusion reaction cross section $\sigma(v_r)$ is given by
\begin{equation}
\sigma(v_r)=\frac{S(v_r)}{\mu v_r^2/2}e^{-v_{\rm G}/v_r}, \label{eq:cross}
\end{equation}
where $v_{\rm G}=\sqrt{2E_{\rm G}/\mu}$, $E_{\rm G} = 2 \mu c^2 (\pi \alpha Z_1 Z_2)^2$ is the Gamow energy, and $\mu=m_im_j/(m_i+m_j)$ is the reduced mass. $S(v_r)$ is the astrophysical S-factor, which is customarily fitted by rational functions of the kinetic energy $E=\mu v_r^2/2$ \cite{Bosch_1992}. For ICF-relevant energies, $\sigma(v_r)$ is dominated by the exponential term in Eq.\ (\ref{eq:cross}), implying that reactants with larger relative velocities possess significantly larger cross sections. However, in a Maxwellian (or near-Maxwellian) plasma, this steep increase in cross section competes with an exponential decay of the distribution function. When $f_i$ and $f_j$ are both Maxwellian, Eq.\ (\ref{eq:1}) reduces to a one-dimensional integral,
\begin{equation}
    \langle\sigma v\rangle=\frac{2-\delta_{ij}}{2}\int v_r\frac{4S(v_r)}{\sqrt{\pi}v_{th}^3} e^{-v_{\rm G}/v_r-v_r^2/v_{th}^2}d v_r,\label{eq:reactivity}
\end{equation}
where $v_{th}=\sqrt{2T/m}$ is the thermal velocity. Neglecting the relatively weak dependence of the prefactors on $v_r$, the integrand in Eq.\ (\ref{eq:reactivity}) reaches its maximum at $v_p=(v_{\rm G}v_{th}^2/2)^{1/3}$, which corresponds to the Gamow peak $E_P=\mu v_r^2/2=(E_{\rm G}T^2/4)^{1/3}$. Given $v_{\rm G}\gg v_{th}$, it follows that $v_p\gg v_{th}$, implying that ions from the high-energy tail of the distribution function, which constitute only a small fraction of the total population, contribute the bulk of the fusion reactivity.

\subsection{Collision frequency}

In general, kinetic effects in collisional plasmas hinge on the scale separation between the tail ions and the thermal bulk, which also underpins SFRE discussed in this work. This separation arises from the velocity dependence of the collision frequency in plasmas. For a test particle $i$ colliding with a stationary background species $j$, the collision frequency $\nu\equiv-\langle d\bm v/dt\rangle/\bm v$ is given by
\begin{equation}
\nu_0(v) =\frac{n_j}{v^3}\Gamma_{ij}\frac{m_i}{\mu}= \frac{n_j Z_i^2 Z_j^2 e^4 \ln \Lambda}{4\pi \epsilon_0^2 m_i \mu v^3}, \label{eq:collision}
\end{equation}
where $\ln \Lambda$ is the Coulomb logarithm. If the background species $j$ possesses some velocity distribution instead of being static, the collision frequency needs to be evaluated from kinetic theory with specific collision operators. Frequently employed is the FP operator, which is defined as 
\begin{eqnarray}
   \mathcal C[f_i] = - \frac{\partial}{\partial \bm v_i} \left[ \langle \Delta \bm  v_i \rangle f_i \right] + \frac{1}{2}  \frac{\partial^2}{\partial \bm v_i \partial \bm v_i} \bm :\left[ \langle \Delta \bm v_i \Delta \bm v_i \rangle f_i \right], \label{eq:FP}
\end{eqnarray}
where $\langle \Delta \bm  v_i \rangle$ and $\langle \Delta \bm v_i \Delta \bm v_i \rangle$ represent the expectation of the first and second moments of the velocity increment per unit time, respectively. These coefficients are expressed in terms of the Rosenbluth potentials \cite{Rosenbluth1957}, where $\langle \Delta \bm  v_i \rangle=\sum_j(\Gamma_{ij}m_i/\mu)\nabla_{\bm v_i}H_j$ and $\langle \Delta \bm v_i \Delta \bm v_i \rangle=\Gamma_{ij}\nabla_{\bm v_i} \nabla_{\bm v_i} G_j $. The potentials are defined by the following integrals:
\begin{eqnarray}
 &&   H_j(\bm{v}) = \int \frac{f_j(\bm{v}')}{|\bm{v}-\bm{v}'|} d\bm{v}',\\ 
 &&G_j(\bm{v}) = \int |\bm{v} - \bm{v}'| f_j(\bm{v}') d\bm{v}'. \label{eq:Gpotential}
\end{eqnarray}
If $f_j$ is a Maxwellian distribution, the potentials $H_j$ and $G_j$ possess analytical solutions that depend only on the velocity magnitude $v_i = |\bm v_i|$. Specifically, $H_j = (n_j/v_i)\text{erf}(x)$, where $x = v_i / v_{th,j}$. Substituting this into the expression for $\langle \Delta \bm v_i \rangle$, the effective collision frequency given by the single background species $j$ is derived as:
\begin{equation}
\nu_M(v) = \nu_0(v)\left[ \text{erf}(x) - x\text{erf}'(x) \right].
\label{eq:collision2}
\end{equation}
For tail ions with $v\gg v_{th,j}$, Eq.\ (\ref{eq:collision2}) naturally reduces to Eq.\ (\ref{eq:collision}). From Eqs.\ (\ref{eq:collision}) and (\ref{eq:collision2}), there is $\nu(v) \propto v^{-3}$. From this relation, it follows that
\begin{equation}
\lambda(v) \approx \frac{v}{\nu(v)} \propto v^4, \label{eq:mfp}
\end{equation}
where $\lambda$ is the mean free path of the test particle. The scaling in Eqs.\ (\ref{eq:collision}) and (\ref{eq:mfp}) provides the physical mechanism for the SFRE. While the thermal bulk ($v \approx v_{th}$) is strongly collisional and constrained by the local fluid velocity field, the tail ions participating in fusion ($v \approx v_p \gg v_{th}$) suffer from collisions far less frequently. Their long mean free paths allow them to decouple from the local fluid and traverse macroscopic velocity gradients, thereby establishing non-Maxwellian tail distributions non-locally. At a typical temperature $T=4$\ keV, for DT fusion there is $v_{\rm G}/v_{th}\approx2.05$, which implies a scale separation of $\lambda$ at about 17.6 times.

\subsection{Tail enhancement by turbulence}

Now we proceed to describe the enhanced tail distribution established by turbulent flows. For simplicity, we only consider a single ion species. The  kinetic equation is
\begin{equation}
    \frac{\partial f}{\partial t}+\bm v\cdot \nabla f+\bm a\cdot \frac{\partial f}{\partial \bm v}=\mathcal C[f], \label{eq:kinetic}
\end{equation}
where $a$ is the acceleration and $\mathcal C[f]$ is the collision operator. In terms of the peculiar velocity $\bm w=\bm v-\bm u$, Eq.\ (\ref{eq:kinetic}) can be rewritten as
\begin{equation}
    \frac{\partial f}{\partial t}+(\bm w+\bm u)\cdot \nabla f+(\bm a-\frac{d \bm u}{dt}-\bm w\cdot\nabla \bm u)\cdot \frac{\partial f}{\partial \bm w}=\mathcal C[f].
\end{equation}
Since the dynamical timescale of the tail ions is much shorter than that of the thermal particles, we can neglect the evolution of the macroscopic flow, and seek a quasi-steady-state solution satisfying
\begin{equation}
    (\bm{w}+\bm{u})\cdot\nabla f - \bm{w}\cdot\nabla \bm{u} \cdot \frac{\partial f}{\partial \bm{w}} = \mathcal C[f]. \label{eq:steady_state}
\end{equation}
Given that the tail particles only constitute a small part of the total distribution, we can  decompose the distribution function into the local Maxwellian $f_M$ and non-Maxwellian perturbations $f_n$. Assuming the flow field is incompressible ($\nabla\cdot\bm{u}=0$) and hence subsonic ($u\ll v_{th}$), $f_n$ can be sorted by orders of $u$. Substituting this expansion into Eq.\ (\ref{eq:steady_state}) and retaining only first-order terms yields the linearized steady-state equation:
\begin{equation}
    \bm{w}\cdot\nabla f_1 - \bm{w}\cdot\nabla \bm{u} \cdot \frac{\partial f_M}{\partial \bm{w}} = \hat{\mathcal{C}}[f_1], \label{eq:1stperturbation}
\end{equation}
where $\hat{\mathcal{C}}$ is the generic linearized collision operator. For a single Fourier mode with wavevector $\bm{k}$ and velocity amplitude $\tilde{\bm{u}}(\bm{k})$, where $\bm{k} \perp \tilde{\bm{u}}$, the spatial gradient transforms as $\nabla \to i\bm{k}$. Without loss of generality, we can assume $\bm k=k \hat {\bm e}_z$ and $\tilde {\bm u}=\tilde {u} \hat {\bm e}_x$, and Eq.\ (\ref{eq:1stperturbation}) can be rewritten as
\begin{equation}
    ikw_z\tilde{f}_1(\bm{k}, \bm{w}) - ikw_z\left[\tilde{{u}}(\bm{k})\cdot \frac{\partial f_M}{\partial w_x}\right] = \hat{\mathcal{C}}[\tilde{f}_1].\label{eq:f1}
\end{equation}
Using $\partial f_M/\partial w_x=-2f_Mw_x/v_{th}^2$, we can express the perturbed distribution $\tilde{f}_1$ via the inverse of the linear operator:
\begin{equation}
    \tilde{f}_1(\bm{k}, \bm{w}) = \left[ ikw_z - \hat{\mathcal{C}} \right]^{-1} \left[ -\frac{2ikw_zw_x}{v_{th}^2} \tilde{u}(\bm{k}) f_M\right]. \label{eq:generalized_f1}
\end{equation}
Equation\ (\ref{eq:generalized_f1}) serves as the generalized formal solution for the first-order steady-state perturbation. The term inside the brackets on the right-hand side acts as an anisotropic driving source, with the tail enhancement arising from the product of $w_z$ and $w_x$. The physical essence of the tail enhancement is encoded within the resolvent operator $[ ikw_z - \hat{\mathcal{C}} ]^{-1}$, which represents the competition between the collisionless free-streaming term $ikw_z$  and collisional isotropization $\hat{\mathcal{C}}$. 

Provided $\hat{\mathcal{C}}$ possesses reflection symmetry in the phase space, Equation (\ref{eq:generalized_f1}) is an odd function of $w_x$,  yielding zero net enhancement in fusion reactivity when reacting with $f_M$ (i.e., $\Sigma[f_1,f_M]=0$).  Consequently, the lowest-order reactivity enhancement is second-order, $\Sigma[f_1,f_1]+\Sigma[f_M,f_2]$, which necessitates computing the second-order perturbation $f_2$.

Expanding Eq.\ (\ref{eq:steady_state}) to the second order in $u$ yields:
\begin{equation}
    \bm{w}\cdot\nabla f_2+\bm u\cdot \nabla f_1 - \bm{w}\cdot\nabla \bm{u} \cdot \frac{\partial f_1}{\partial \bm{w}} = \hat{\mathcal{C}}[f_2], \label{eq:2ndperturbation}
\end{equation}
which introduces the nonlinear interaction between $u$ and $f_1$. In Fourier space,
\begin{eqnarray}
   i(\bm{k}\cdot\bm{w})\tilde{f}_2(\bm{k}, \bm{w})-&&\hat{\mathcal{C}}[\tilde{f}_2] =
  \sum_{\bm{k}_1+\bm{k}_2=\bm{k}} \bigg[ i(\tilde{\bm{u}}(\bm{k}_1)\cdot \bm{k}_2)\tilde{f}_1(\bm{k}_2, \bm{w})\nonumber \\ 
  &&  - i(\bm{w}\cdot\bm{k}_1)\tilde{\bm{u}}(\bm{k}_1)\cdot \frac{\partial \tilde{f}_1(\bm{k}_2, \bm{w})}{\partial \bm{w}}\bigg] . \label{eq:2ndfourier}
\end{eqnarray}
Here, we no longer assume fixed directions for $\bm{k}$ and $\tilde{\bm{u}}$, since different Fourier components over the whole space are involved. Equation (\ref{eq:2ndfourier}) gives a lengthy expression for $\tilde f_2$:
\begin{eqnarray}
   && \tilde{f}_2(\bm{k}, \bm{w}) = \left[ i(\bm{k}\cdot\bm{w}) - \hat{\mathcal{C}} \right]^{-1} \sum_{\bm{k}_1+\bm{k}_2=\bm{k}}i\bigg[ (\bm{w}\cdot\bm{k}_1) \nonumber \\ &&\tilde{\bm{u}}(\bm{k}_1)\cdot \frac{\partial \tilde{f}_1(\bm{k}_2, \bm{w})}{\partial \bm{w}} - (\tilde{\bm{u}}(\bm{k}_1)\cdot \bm{k}_2)\tilde{f}_1(\bm{k}_2, \bm{w}) \bigg]. \label{eq:generalized_f2}
\end{eqnarray}
Now, to the second order, the reactivity enhancement factor can be written as 
\begin{equation}
    \Phi[f,f]=1+\frac{\Sigma[f_1,f_1]+\Sigma[f_2,f_M]}{\Sigma[f_M,f_M]}.
\end{equation}
To explicitly evaluate Eqs.\ (\ref{eq:generalized_f1}) and (\ref{eq:generalized_f2}), one must specify the mathematical structure of $\hat{\mathcal{C}}$. While the FP operator in Eq.\ (\ref{eq:FP}) is largely accurate physically, it precludes an analytical solution of Eqs.\ (\ref{eq:generalized_f1}) and (\ref{eq:generalized_f2}). Alternatively, FF employed a modified BGK operator \cite{BGK1954},
\begin{equation}
    \mathcal{C}[f_i]=-\nu (w)(f_j-f_M),
\end{equation}
so $\hat{\mathcal{C}}$ collapses into a simple scalar multiplication by $-\nu(w)$, and the inverse operator yields a straightforward algebraic fraction, recovering the exact analytical form derived by FF \cite{fetsch2025analytical}:
\begin{equation}
    \tilde{f}_1(\bm{k}, \bm{w}) = -\frac{2}{v_{th}^2}\frac{ikw_zw_x\tilde{u}(\bm{k}) f_M}{ ikw_z +\nu(w)}. \label{eq:BGKsol}
\end{equation}
Nevertheless, Eq.\ (\ref{eq:generalized_f2}) still involves an infinite summation over $k$. FF proposed that for $\bm k\neq0$, the spatial average of the enhancement $\langle \Sigma[f_M,\tilde f_2(\bm k)]\rangle$ vanishes, and thus only $\tilde f_2(0,\bm w)$ needs to be considered. In this case, only $\tilde {\bm u}(\bm k_1)$ and $\tilde f_1(\bm k_2,\bm w)$ with $\bm k_2=-\bm k_1$ contribute, which greatly simplifies the computation and enables an analytical expression of $\tilde f_2$. 

Despite the significance of FF's explicit formula, several limitations clearly exist. First, the exclusion of $\tilde f_2(\bm k\neq0,\bm w)$ potentially overlooks real physics. Even if the spatial average is zero, local enhancements can trigger a positive feedback and result in a net enhancement considering the real temporal evolution. Second, the results are derived using a perturbation theory assuming small values of $u$, which must fail in supersonic ($u>v_{th}$) turbulence. Third and most importantly, in spite of the clarity of the solution, BGK operator is an oversimplification on collision. Compared with the FP operator Eq.\ (\ref{eq:FP}), the BGK operator only mimics the first term of FP operator, i.e., the slowing-down term, while the second term representing velocity diffusion and pitch-angle scattering is completely neglected. The latter acts as an isotropization force, which tends to ``dilute'' the high-energy tail, and neglecting it likely leads to an overestimated $\Phi$. Of course, it is impossible to incorporate all these effects in an analytical expression for $\Phi$, especially when the FP operator renders Eqs.\ (\ref{eq:generalized_f1}) and (\ref{eq:generalized_f2}) intractable. Henceforth, in the next Section we will present numerical calculations of the steady-state tail distribution.

\section{Steady state tail distribution}\label{sec:tail}
\subsection{Simulation settings}
To investigate the tail distribution established by the shear flow, we numerically solve for the steady-state distribution and calculate the resulting SFRE, using both BGK and FP operators. We consider an isothermal plasma with a single ion species. For simplicity and tractability, a single-mode shear flow is prescribed as a rigid background:
\begin{equation}
    \bm{u}(z)=u_0\sin(kz)\hat{\bm e}_x. \label{eq:shearflow}
\end{equation}
Now we seek a solution of $f$ satisfying the steady state equation, Eq.\ (\ref{eq:steady_state}). The simulation is performed in a 1D2V phase space $(z, v_x, v_z)$. The spatial domain covers one wavelength $L = 2\pi/k$ with periodic boundary conditions, while the velocity domain is truncated at $\pm 8v_{th}$ in both directions to sufficiently capture the tail distribution.

For the BGK operator, Eq.\ (\ref{eq:steady_state}) is solved with the method of characteristics, where the distribution is propagated ballistically along $v_z$, and collision acts as an attenuation factor:
\begin{eqnarray}
f(v,z+\Delta z)  = && f(v,z) e^{-\nu_B(w) \Delta t} \nonumber \\
&&+ f_M(w) (1 - e^{-\nu_B(w) \Delta t}), 
\end{eqnarray}
where $\Delta t=\Delta z/v_z$, and the local Maxwellian is computed from the prescribed shear flow via $f_M(w)=f_M(|\bm v-\bm u(z)|)$. In accordance with FF, the collision frequency here is modified to include a constant electron drag term:
\begin{equation}
    \nu_B(w)=\nu_M(w)+\nu_0(v_{th})\sqrt{ \frac{m_e}{m_i}},\label{eq:freqBGK}
\end{equation}
where $\nu_0$ and $\nu_M$ are defined by Eqs.\ (\ref{eq:collision}) and (\ref{eq:collision2}). This additional term results from the fact that extremely fast ions have a larger collision frequency with electrons than thermal ions. The Coulomb logarithm $\ln \Lambda$ in $\nu$ is derived as a constant from the empirical formula
\begin{equation}
\ln \Lambda_{ii} = 30 - \ln \left(\frac{ Z_1 Z_2 (\mu / m_p) n_i[\rm m^{-3}]^{1/2}}{T_i[\rm eV]^{3/2}} \right).    
\end{equation}
Since $\nu_{B}(w)$ is not a constant, the BGK operator no longer conserves the total density and momentum. After each iteration, we renormalize the distribution function to conserve the local density. The final solution is obtained after 30 iterations. 

For the FP operator, with Eq.\ (\ref{eq:kinetic}) the system is evolved in time until a steady state is reached. For the purpose of maintaining numerical stability, Equation (\ref{eq:kinetic}) is reconstructed with the renormalized flux form of the FP operator:
\begin{equation}
    \frac{\partial f}{\partial t} + v_z \frac{\partial f}{\partial z} = \nabla_v \cdot \left[ f_M \mathbf{D}(\bm w) \cdot \nabla_v \left( \frac{f}{f_M} \right) \right],
\end{equation}
where $f_M$ is still computed from the local flow field. The diffusion tensor $\mathbf{D}(\bm w)$ is derived from the second Rosenbluth potential $G_j$:
\begin{equation}
    \mathbf{D}(\bm w) = \frac{1}{2} \langle \Delta \bm w \Delta \bm w \rangle = \frac{1}{2} \Gamma_{ij} \nabla_{w} \nabla_{w} G_j. \label{eq:Dtensor}
\end{equation}
To align with the modified BGK collision frequency in Eq.\ (\ref{eq:freqBGK}), $\Gamma_{ij}$  in Eq.\ (\ref{eq:Dtensor}) is modified in the simulation to contain the electron drag term.

The kinetic equation is solved with an operator splitting scheme. In each time step $\Delta t$, the evolution is decomposed into two stages:
\begin{equation}
    f(t + \Delta t) = \mathcal{L}_{FP}(\Delta t) \mathcal{L}_{adv}(\Delta t) f(t).
\end{equation}
The convection term $v_z \partial_z f$ is solved using a Fourier spectral method. By transforming the distribution function into the Fourier domain, the advection is computed as a phase shift: $\hat{f}(k_z, \bm v, t+\Delta t) = \hat{f}(k_z, \bm v, t) e^{-i k_z v_z \Delta t}$. This approach is dissipation-free and dispersion-free, which preserves the tail distribution without the artificial smoothing common in finite-difference methods. The velocity-space diffusion is evolved using an alternating direction implicit (ADI) scheme. The 2D diffusion operator is split into two 1D sub-steps ($v_x$ and $v_z$). Each sub-step involves solving a tridiagonal system, which we implement with an optimized, vectorized Thomas algorithm \cite{Press2007}. To handle the non-diagonal elements of the diffusion tensor $\mathbf{D}(\bm w)$, the cross-derivative terms are treated explicitly within the ADI cycle. This hybrid implicit-explicit approach ensures unconditional stability for the diagonal diffusion, allowing for a large simulation time step to rapidly reach the steady state.

Before presenting the simulation results, it is important to note the dimensional limitations of our framework. While a 1D2V simulation is sufficient and self-consistent for the BGK model, it inherently underestimates the efficiency of pitch-angle scattering in the Fokker-Planck (FP) model. In a full 3V velocity space, the increased rotational degrees of freedom enhance the isotropization of fast ions. Consequently, the reactivity enhancements predicted by the FP operator in this work should be regarded as an upper bound of the actual value.

\subsection{A fiducial case}
A series of simulations is conducted to investigate how the different operators act on the shear flow across a range of parameters. For illustration, we first present the results of a fiducial case with $T=4$ keV, $\rho=150$ g/cm$^3$ and a wave vector $k=2\pi/0.56\ \mu\rm m^{-1}$. This flow wavelength $L=0.56\ \mu$m is relatively small compared to typical turbulence scales in realistic scenarios but features a larger SFRE. The flow amplitude is set to a supersonic value $u_0=\sqrt{3/2}v_{th}$, which means the TKE equals 1/4 of the thermal energy of the system. The plasma is modeled as a single-species  D-T ion mixture with an average mass $m=2.5m_p$.

\subsubsection{Phase space distribution}

\begin{figure}[b]
\includegraphics[width=0.98 \columnwidth]{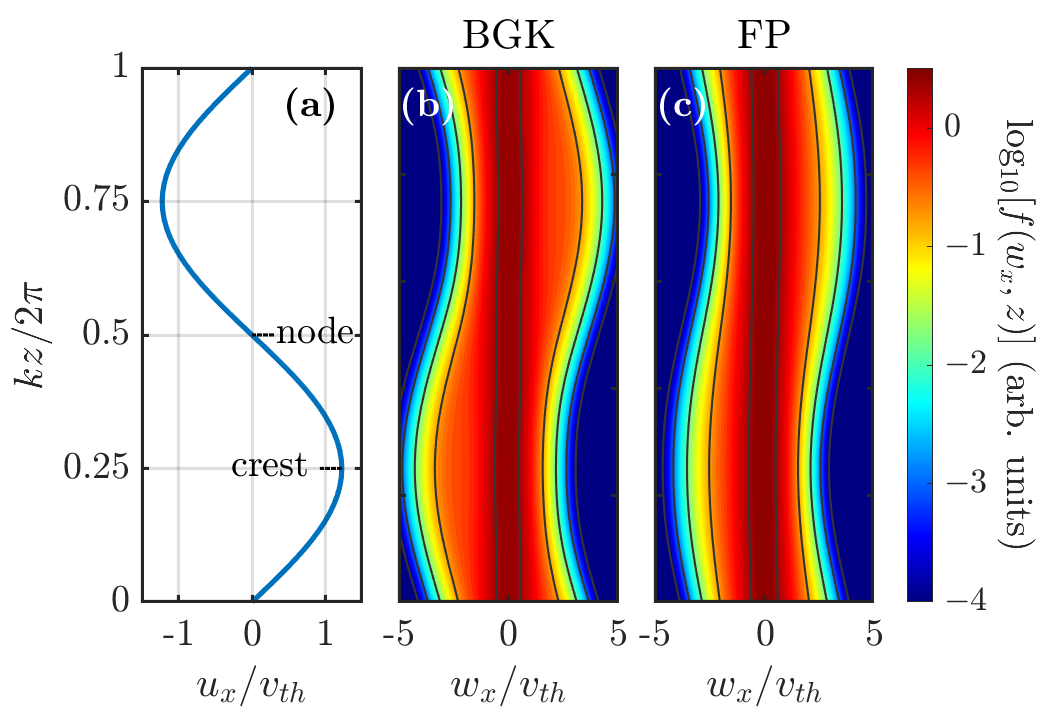}
\caption{\label{fig:zwxspace}(a) Spatial profile of the prescribed shear flow $u_x(z)$ across the simulation domain. (b)-(c) Steady-state ion distributions $f(w_x, z)$ integrated over $w_z$, calculated by the BGK and FP operators, respectively.}
\end{figure}

For both the BGK and FP operators, the simulation reveals significant distortions in the phase space distribution. To draw a general picture, in Fig.\ \ref{fig:zwxspace} we display the spatial profile of the sinusoidal shear flow $u_x$ and the corresponding steady-state distribution $f(w_x,z)$ calculated by both operators. The non-Maxwellian distortions in $w_x$ clearly demonstrate the non-local transport of the tail ions, where a $z$-direction tail profile is formed in antiphase with the flow profile. At the flow crest ($kz=\pi/2$), a biased tail forms in the $w_x<0$ region, which results from the fact that tail particles originating from elsewhere possess a $\langle v_x\rangle<u_0$. When arriving at the crest, they manifest themselves as a biased tail with $\langle w_x\rangle=\langle v_x-u_0\rangle<0$.

While the overall morphology of $f(w_x, z)$ is similar for both operators, the FP operator produces a notably narrower distribution in $w_x$. This suggests that the energy diffusion and pitch-angle scattering inherent in the FP model provides a more efficient dissipation mechanism, scattering tail particles into other velocity dimensions and thus suppressing the buildup of extreme $w_x$ anisotropy compared to the BGK  model.

\begin{figure}[b]
\includegraphics[width=0.98 \columnwidth]{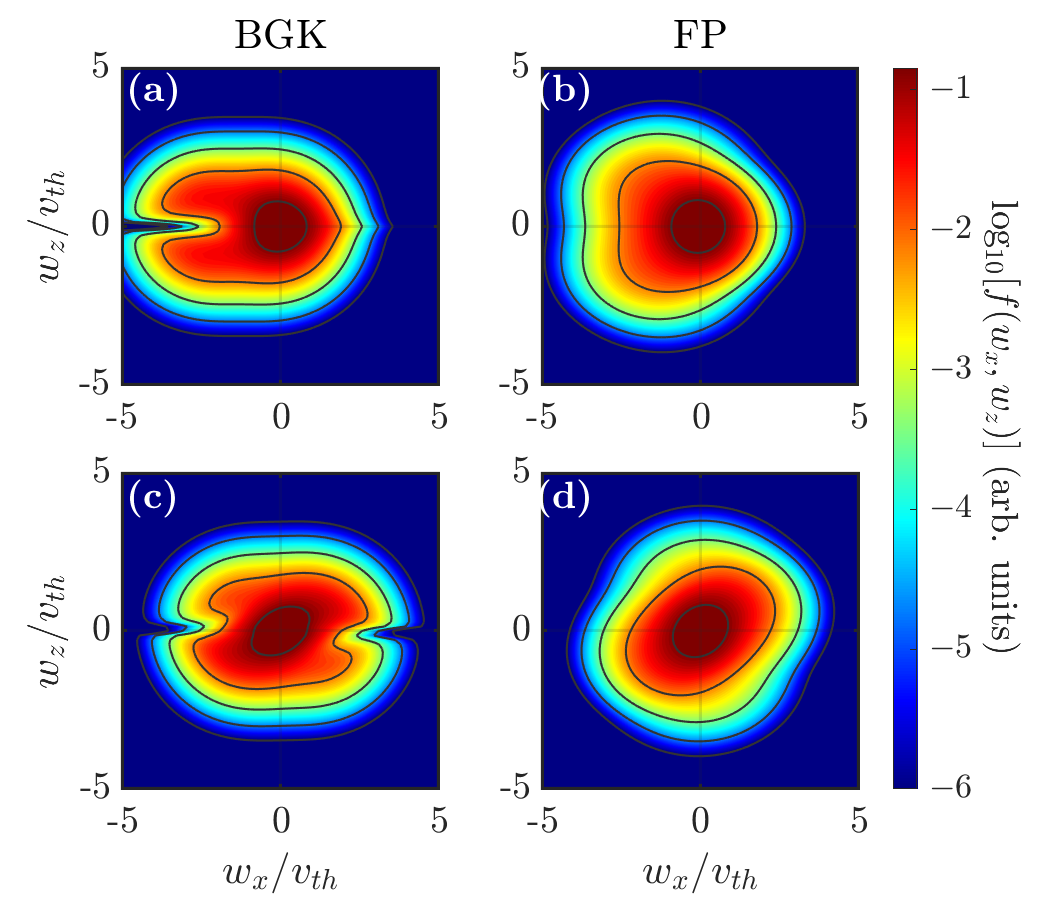}
\caption{\label{fig:pspace} Local steady-state ion distributions $f(w_x,w_z)$, at the flow crest ($kz=\pi/2$, top row) and node ($kz=\pi$, bottom row) calculated by the BGK (left) and FP (right) operators respectively.}
\end{figure}

To further illustrate the structure of the tails, Figure\ \ref{fig:pspace} displays the steady-state distributions $f(w_x,w_z)$ extracted at the flow crest ($kz=\pi/2$) and the node ($kz=\pi$). Generally, the enhanced distribution possesses  a correlation between their $w_z$ and their spatial \textit{origins} $z_o$, where particles with $w_z>0$ originate from upstream regions ($z_o<z$) and vice versa.  This is strictly true for the BGK operator (if we neglect the periodic boundary conditions) and roughly true for the FP operator, with which particles are scattered in the phase space. As shown in Fig.\ \ref{fig:pspace}(a) and (b), the distribution at the flow crest is symmetric with respect to $w_z$.  This stems from the symmetry of the flow about $kz=\pi/2$, and particles from upstream and downstream make the same contribution in the $w_x$ distortion. In contrast, at the flow node, the distorted phase space distribution exhibits central symmetry about the origin (0,0), as seen in Fig.\ \ref{fig:pspace}(c) and (d). This reflects the anti-symmetric nature of the velocity gradient at the node. Particles with $w_z>0$ bring a $\langle w_x\rangle>0$ from upstream, and vice versa, causing the distribution to stretch along the $w_x\sim w_z$ diagonal.

It is noteworthy that in the BGK results, Fig.\ \ref{fig:pspace}(a) and (c), a sharp ``trough'' in the distribution function exists at the line of $w_z=0$. This directly results from the feature of the BGK operator, where particles propagate in space ballistically without rotating and accelerating. This means particles with $w_z=0$ are always constrained on a line with $z=$ const. In consequence, while the tail distribution is enhanced elsewhere, the distribution at $w_z=0$ always remains Maxwellian. 

This characteristic is what distinguishes the two operators in physics. While the BGK operator persistently retains the velocity of the tail particles, the FP operator scatters them in the velocity space. In consequence, the ``trough'' is filled by particles scattered from other regions of the phase space, leading to a more azimuthally diffused and smooth distribution.

\subsubsection{Reactivity Enhancement}

In the context of fusion reactivity, what really matters is the distribution of the relative velocity magnitude $w_r$ between reacting ion pairs. For a given local distribution $f(\bm w)$, the distribution of the relative velocity vector $\bm w_r$  is defined by the self-convolution
\begin{equation}
F(\bm w_r) = \int  f(\bm w) f(\bm w - \bm w_r) d \bm w, \label{eq:conv}
\end{equation}
and the one-dimensional distribution $f(w_r)$ is subsequently derived through an angular average:
\begin{equation}
    f(w_r) = \frac{1}{2\pi} \int_{0}^{2\pi} F(w_r \cos\theta, w_r \sin\theta) d\theta. \label{eq:wr}
\end{equation}

\begin{figure}[b]
\includegraphics[width=0.98 \columnwidth]{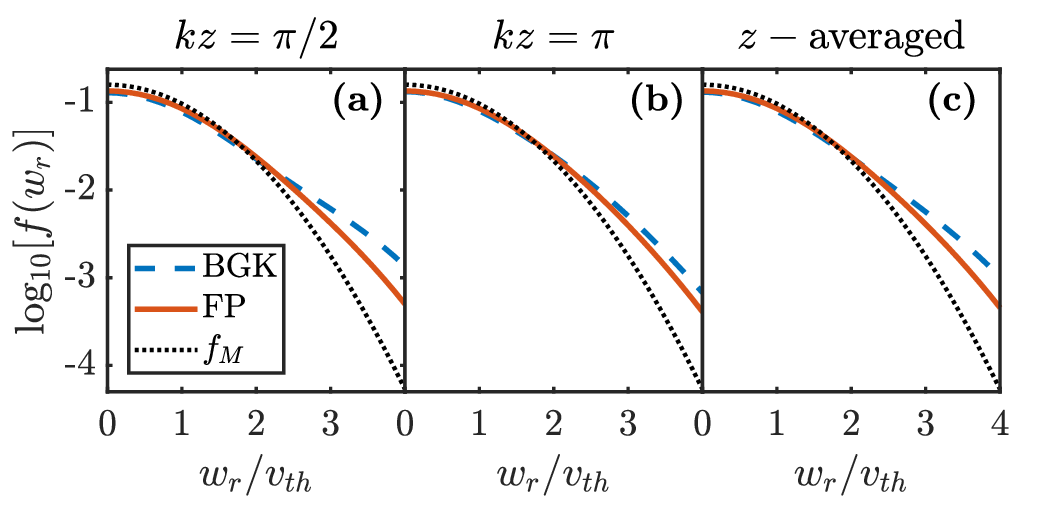}
\caption{\label{fig:1ddis} The distribution $f(w_r)$ of the relative velocity, evaluated at (a), $kz=\pi/2$, (b), $kz=\pi$ and (c), averaged across all $z$. The distributions calculated by the BGK (blue, dashed) and FP (red) operators are compared with the initial Maxwellian (black, dotted).}
\end{figure}

Figure \ref{fig:1ddis} displays the resulting 1D relative velocity distributions calculated in the $w_x-w_z$ phase space. Both the BGK and FP operators predict a significant non-Maxwellian enhancement in the high-energy tail. However, the BGK operator generally yields a more pronounced tail than the FP operator. This is a direct consequence of the more ``structured'' phase space distortions observed in the BGK case [see Fig.\ \ref{fig:pspace}(a) and (c)], which are less mitigated by velocity-space diffusion compared to the FP model.

Specifically, the tail enhancement is notably stronger at the flow crest ($kz=\pi/2$) than at the node ($kz=\pi$), particularly for the BGK operator. Physically, this is due to the formation of a long, biased tail in the negative $w_x$ region at the crest [see Fig.\ \ref{fig:pspace}(a) and (b)]. In the convolution integral Eq.\ (\ref{eq:conv}), this tail yields a relatively large product with the core Maxwellian.

\begin{figure}[b]
\includegraphics[width=0.98 \columnwidth]{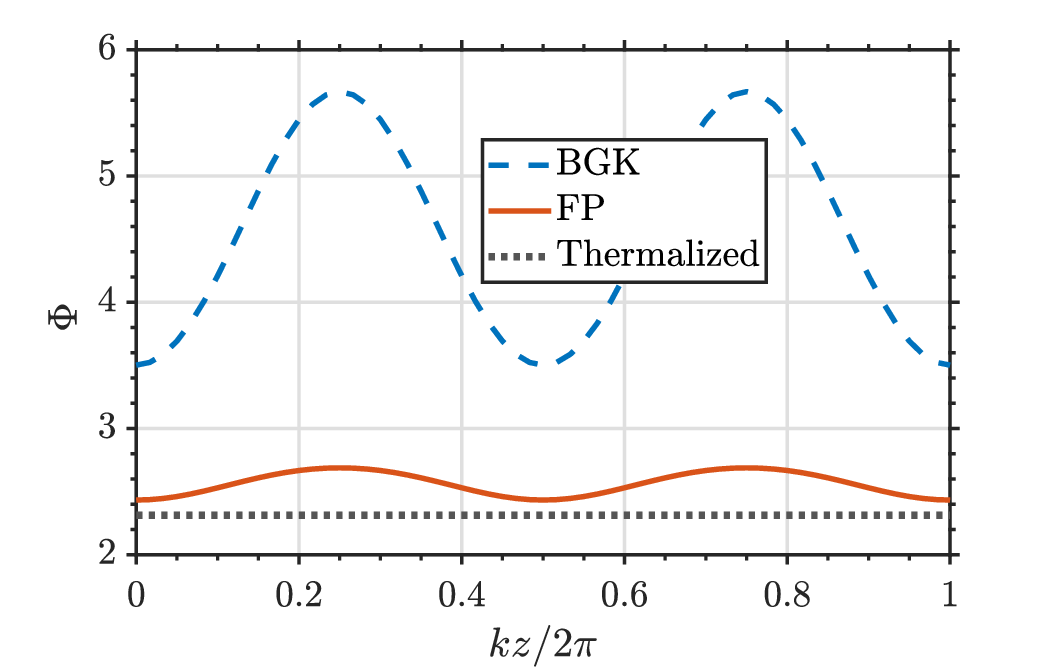}
\caption{\label{fig:Enhance}The fusion enhancement factor $\Phi$ as a function of $z$ evaluated with the BGK (blue, dashed) and FP (red) operators. For comparison, a black dotted line with all the TKE uniformly thermalized into a Maxwellian ($T_{\rm eff}=5$ keV) is also drawn. }
\end{figure}

Clearly, the enhanced tail distribution will lead to a larger fusion reactivity. Although our simulations are performed in a 1D2V framework, the reactivity is calculated by assuming a Maxwellian distribution in the third velocity dimension ($v_y$) and performing the full 3V integral of Eq.\ (\ref{eq:1}). Figure \ref{fig:Enhance} compares the fusion enhancement factor $\Phi[f]=\Sigma[f,f]/\Sigma[f_M,f_M]$ evaluated by both operators. For realistic consideration, we also calculate the case where all the TKE is fully transformed into thermal energy of ions and electrons and forms a new Maxwellian with a higher temperature, which corresponds to $T=5$ keV in this case. 

For both models, $\Phi(z)$ exhibits a cosinoidal shape corresponding to $-\cos (2kz)$, which results from the fact that tail enhancement is reflection-symmetric with respect to the sign of the flow velocity.   Both operators predict a $\Phi$ larger than $\Phi_{th}=2.31$, which means partitioning energy into TKE can, in principle, boost fusion reactivity in this case. However, although the BGK operator predicts a $\langle \Phi\rangle$ up to 4.58, the FP operator only gives a modest value of 2.56, which is slightly above the thermalized benchmark $\Phi_{th}$. As noted previously, the 1D2V FP model inherently underestimates the isotropization rate. Given that the true 3V pitch-angle scattering would further isotropize these non-Maxwellian tails, the actual kinetic enhancement in a realistic counterpart is likely even closer to, or potentially lower than, the thermalized limit.

\subsection{Parameter scanning}

Now we proceed to display results of cases with different parameters. Dimensionally, the tail distribution enhancement only relies on two dimensionless parameters,  $kv_{th}/\nu_0$ and $u_0/v_{th}$; when calculating fusion reactivity, the cross section is a function of the absolute value of the relative velocity, so $v_{th}$ or $T$ acts as a third parameter in determining the enhancement factor $\Phi$. 

\begin{figure}[b]
\includegraphics[width=0.98 \columnwidth]{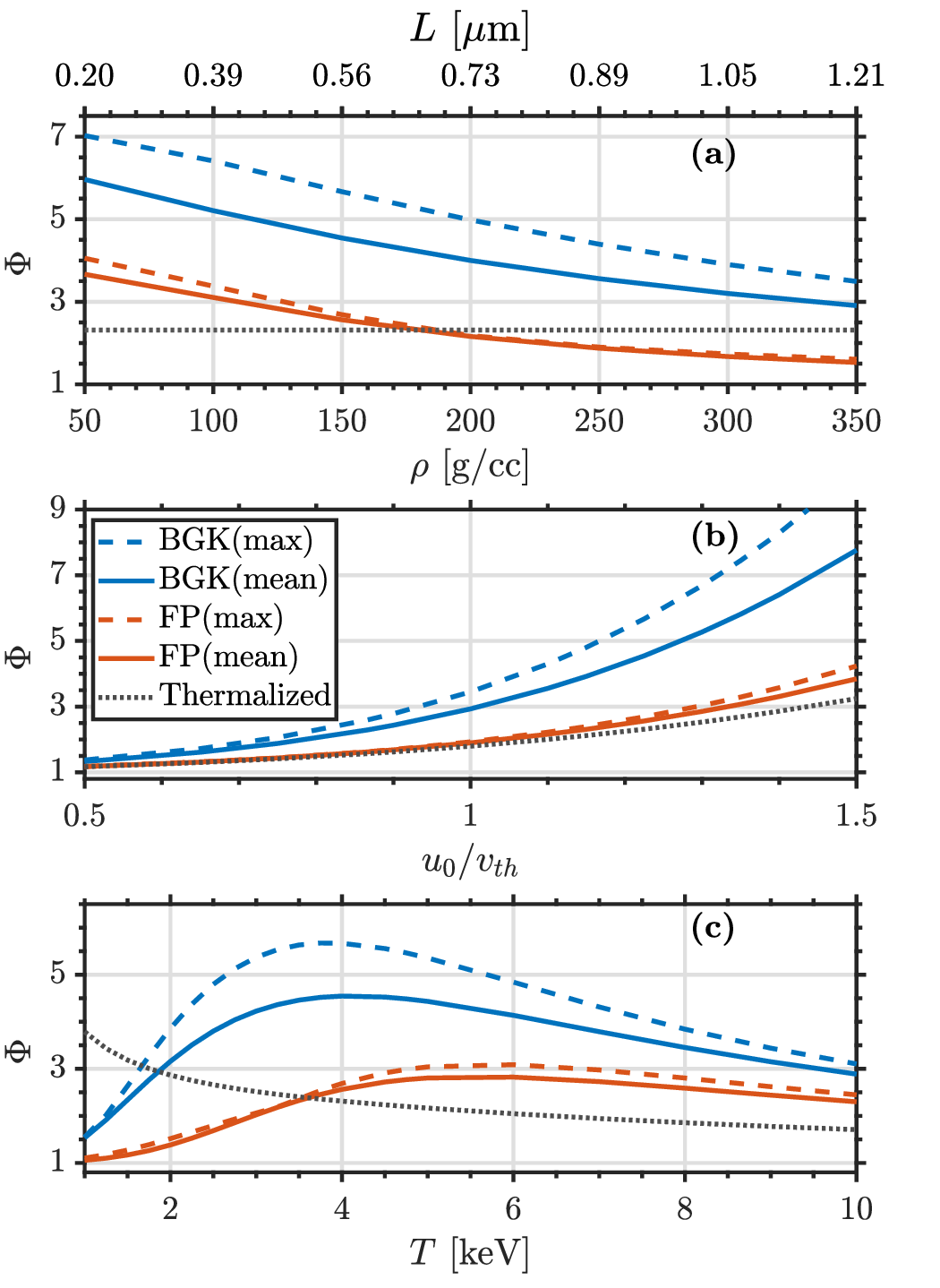}
\caption{\label{fig:scan}The reactivity enhancement factor $\Phi$ in simulations with different parameters, with one parameter shifted each time compared to the fiducial case: (a) the density $\rho$. Since $\rho$ is only related to $\nu_0$, at the top we also mark the equivalent scenario (constant $kv_{th}/\nu_0$) with $\rho=150$ g/cc fixed but the wavelength $L=2\pi/k$ shifted. (b) the flow amplitude $u_0$,  and (c) the ion temperature $T$. In each panel, we plot the BGK (blue) and FP (red) results, with both the maximal $\Phi(z)$ (dashed) and the average $\langle\Phi (z)\rangle$. The dashed gray line represents $\Phi_{th}$ with the TKE thermalized.}
\end{figure}

Figure\ \ref{fig:scan} shows the results of simulations with one parameter shifted each time compared to the fiducial case. In Panel (a) we vary the density $\rho$,  which only shifts the collision frequency of the system. In principle, this is equivalent to varying $k=2\pi/L$ with $\rho$ fixed. All $\Phi$ are monotonically decreasing function of $\rho$ and also $L$ except for $\Phi_{th}$, which is merely a constant under different $\rho$. As the collision frequency increases with the density, the mean free path $\lambda$ of the tail particles reduces (or equivalently $L$ increases compared to $\lambda$). This leads to reduction of the tail enhancement and henceforth $\Phi$. In consequence, SFRE goes through a shift from more beneficial than thermalizing ($\Phi>\Phi_{th}$) to less beneficial ($1<\Phi<\Phi_{th}$). In Panel (b) $u_0/v_{th}$ is shifted, where all $\Phi(u_0)$ are simply monotonically increasing functions of $u_0$, which is not surprising. Generally, it is $kv_{th}/\nu_0$ that decides if SFRE is more beneficial than thermalizing or not, while $u_0/v_{th}$ only decides the magnitude of the effect. 

In Figure\ \ref{fig:scan}(c), we examine the impact of the ion temperature $T$ on the enhancement factor. The thermalized benchmark $\Phi_{th}$ is a monotonically decreasing function of $T$. This originates from the fact that $v_p/v_{th}\propto T^{-1/6}$, which implies the thermal bulk makes more contribution to the total fusion reactivity as $T$ increases. On one hand, this renders the reactivity less sensitive to temperature increments, i.e., $\partial^2\langle\sigma v\rangle/\partial T^2<0$, resulting in the monotonically decreasing $\Phi_{th}$. On the other hand, this also means the reactivity becomes less sensitive to the tail distortion. Consequently, the SFRE factor $\Phi(T)$ exhibits a non-monotonic dependence on $T$, which results from a competition between the increasing mean free path and the diminishing relative contribution of tail ions. This competition gives rise to an optimal temperature for leveraging SFRE.
\begin{figure}[b]
\includegraphics[width=0.98 \columnwidth]{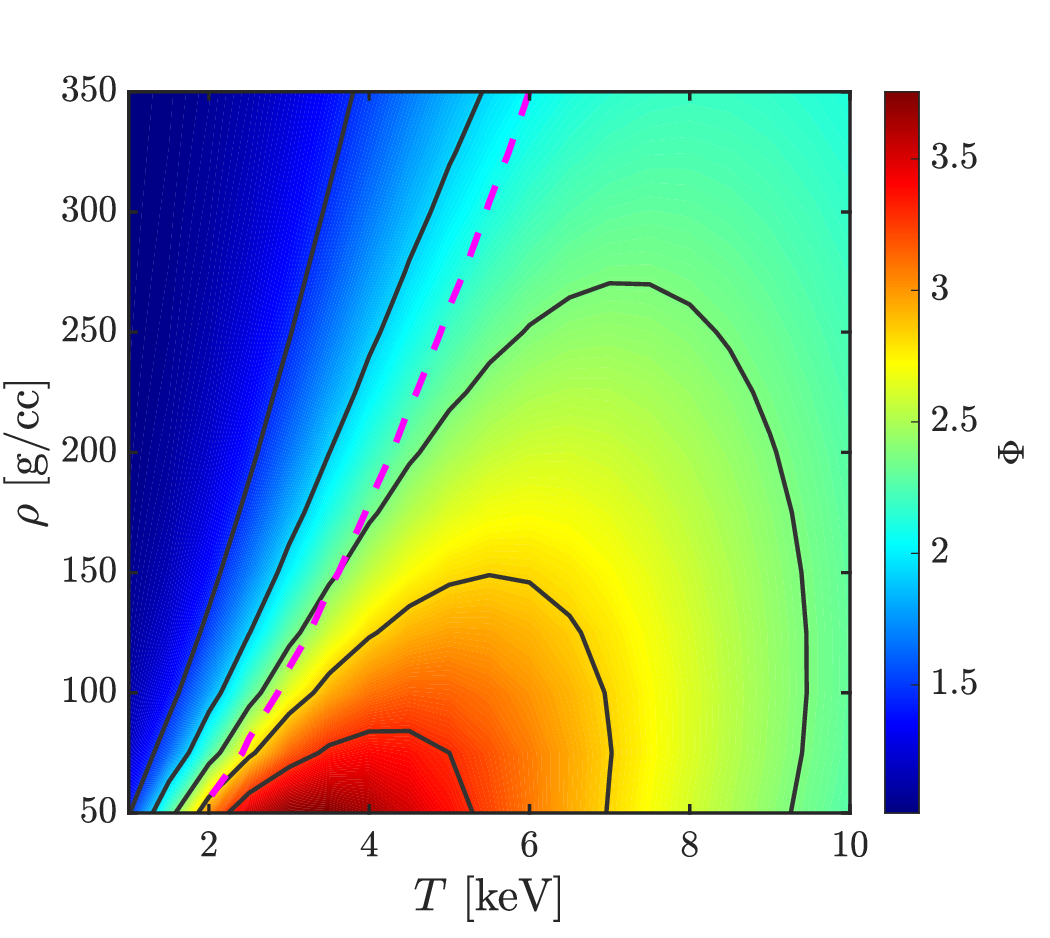}
\caption{\label{fig:scan2d}The averaged $\Phi$ calculated by the FP operator, with a parameter scanning over $T=1\sim10$ keV and $\rho=50\sim350$ g/cc. The shear flow is constantly driven at $k=2\pi/0.56\ \mu\text{m}^{-1}$. The magenta dashed line represents the boundary deciding whether driving the shear flow is more beneficial to reactivity than thermalizing TKE, where at the right side $\Phi>\Phi_{th}$.}
\end{figure}

To provide a more convenient overview, in Fig.\ \ref{fig:scan2d} we show the FP averaged $\langle\Phi\rangle$ in the $\rho-T$ parameter space, where the shear flow is driven at a constant wavenumber $k=2\pi/0.56\ \mu\text{m}^{-1}$. The SFRE is most beneficial at the area of low densities and medial temperatures, where a local maximum $\langle\Phi\rangle\approx3.75$ is found at $3.5$ keV for 50 g/cc. A magenta dashed line represents the boundary where $\langle\Phi\rangle=\Phi_{th}$, while at the right side of the line there is always $\Phi>\Phi_{th}$. As $\rho$ increases, the optimal $T$ also increases to balance the mean free path. 

We also note again that the shear flow wavelength $L=0.56\ \mu$m is small for some parameters. In a plasma, the dissipation timescale of a turbulence with the wavenumber $k$ is given by $\tau_v=1/\eta k^2$, where $\eta\approx 1.8 v_{th}^2/\nu_0(v_{th})$ is the kinematic viscosity. For the most viscous case in our parameter space $T=10$ keV and $\rho=50$ g/cc, the dissipation timescale is $\tau_v\approx0.08$ ps, which is rather small compared to the total ICF timescale and even smaller than the fast particle crossing time $\tau_t\equiv L/v_p\approx0.25$\ ps. Turbulence at this scale dissipates before the tail distribution completely builds up and yields a much smaller $\Phi$ than our prediction. For our fiducial case, $\tau_v\approx1.8$ ps and $\tau_t\approx0.34$ ps, so the time is sufficiently long for SFRE to make a difference before dissipating. However, $\tau_v$ is still small compared to the total burn duration, so turbulence at this scale needs to be triggered at a late time or cascade from turbulence at larger scales to make the best of SFRE. Instead, in a case with $L=3\ \mu$m, $\rho=150$\ g/cc, and $T=4$\ keV, the dissipation time is increased to $\tau_v\approx51.8$\ ps, which is almost comparable to the burn duration. This means turbulence triggered at this scale by hydrodynamic instabilities can result in SFRE directly, although the effect is very modest, where $\langle\Phi\rangle\approx1.087<\Phi_{th}$.  In general, turbulence with larger wavelength will sustain for a longer timescale, but contributes a smaller SFRE.

\subsection{Phase reversal in $\Phi(z)$}
\begin{figure}[b]
\includegraphics[width=0.98 \columnwidth]{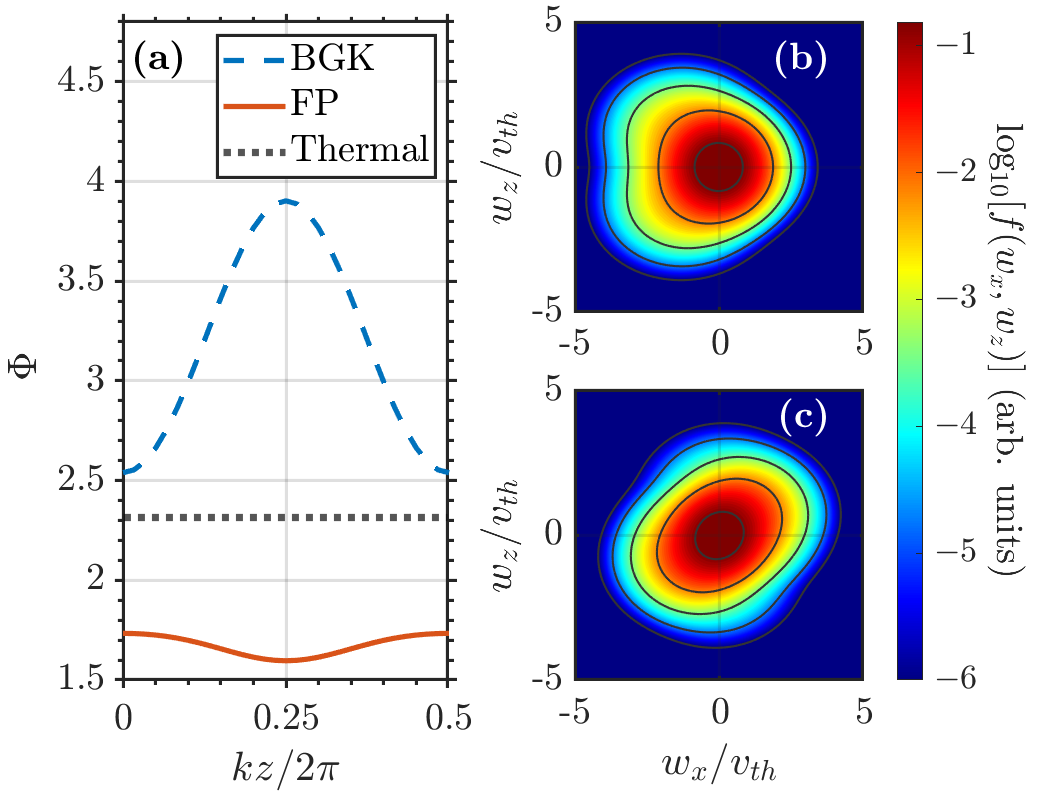}
\caption{\label{fig:300gcc}The results of the case with 300 g/cc and $T=4$ keV. (a) The enhancement factor $\Phi$. Only half of the domain is displayed, while the other half is repetitive. (b) The FP phase space distribution at $kz=\pi/2$. (c) The FP phase space distribution at $kz=\pi$.}
\end{figure}

In Fig.\ \ref{fig:scan}(a) and (c), the FP maximum  and averaged $\Phi$ seem to have a cross at a certain point, which indicates an abnormal constant $\Phi(z)$ across the domain. It turns out that as the collision frequency increases, a spatial phase reversal occurs in the reactivity enhancement factor function $\Phi(z)$. Figure \ref{fig:300gcc}(a) shows $\Phi(z)$ extracted from the case with $\rho=300$ g/cc, where only half of the domain is displayed. It can be seen that $\Phi(z)$ exhibits a $\cos(2kz)$ profile for the FP operator, while the BGK curve still exhibits a $-\cos(2kz)$ shape observed in the $150$ g/cc case. This shift indicates that for the FP operator, the fusion boost is now more pronounced at the flow nodes than at the crests, which is a complete reversal of the low-density behavior.

Figure \ref{fig:300gcc}(b) and (c) show the FP phase space distributions  from the 300 g/cc case, extracted at the crest $kz=\pi/2$ and node $kz=\pi$ respectively. Compared to the 150 g/cc results [Fig.\ \ref{fig:pspace}(b) and (d)], the biased tail in the $w_x<0$ region at the crest shrinks  as the density doubles (most clearly seen from the shape of the second contour), while at the node the shrink is more subtle. The reason is two-fold: at the flow crest $kz=\pi/2$, the tail with the largest $|\bm w|$ originates from the flow trough $kz=3\pi/2$, where the velocity difference is maximal $\Delta u=2u_0$, and the mean free path required is $\lambda\gtrsim L/2$. Meanwhile, at the flow node $kz=\pi$, the tail with the largest $|\bm w|$ originates from both the adjacent crest $kz=\pi/2$ and the trough $kz=3\pi/2$, where $\Delta u=\pm u_0$ and the free path is halved to $\lambda\gtrsim L/4$. As the density and the collision frequency increase, the mean free path for the tail particles falls from $\lambda_p=v_p/\nu_M(v_p)\approx0.45\ \mu\text{m}>L/2$ to $L/2>0.24\ \mu\text{m}>L/4$, so the tail is screened at the crest, while the tail at the node survives. Furthermore, at the node, the tail particles from upstream and downstream deposit in opposite quadrants of the $w_x-w_z$ phase space, which maximizes the relative velocity $w_r$ between ion pairs. In consequence, a larger tail distribution in $f(w_r)$ is formed at $kz=\pi$, yielding a local maximum in $\Phi(z)$. If the collision frequency is improved further, the tail distribution will depend on only the local flow velocity gradient, maintaining $\Phi(z)$ maximized at the node. 

The FP operator, by incorporating pitch-angle scattering, is more sensitive to these collisional screening effects than the BGK model, leading to an earlier phase reversal. Supplemental simulations confirm that at sufficiently high collision frequencies, the BGK model eventually undergoes a similar phase transition.

\section{PIC simulations}\label{sec:pic}

\subsection{Simulation settings}
Now, to provide a more straightforward understanding on SFRE, we conduct 1D3V PIC simulations with the hybrid version of a PIC code LAPINS \cite{Wu2023}. The hybrid code treats ions as kinetic macro-particles while electrons are modeled as a massless fluid neutralizing the ion charge \cite{10.1063/5.0146529}. Collisions in a cell are modeled as binary collisions between randomly paired particles \cite{TAKIZUKA_1977,Nanbu_1998}. Different from the simplified kinetic solver used in Sec.\ \ref{sec:tail}, LAPINS calculates the Coulomb logarithm  from an improved formula dependent on particle velocities \cite{Perez2012}. Fusion reactions are self-consistently incorporated in LAPINS with a stochastic Monte-Carlo sampling algorithm to determine the occurrence of fusion reactions and the generation of reaction products. This enables a self-consistent evaluation of SFRE in a turbulent hot spot.

In 1D3V simulations, turbulence is still modeled as a single-mode sinusoidal shear flow like in Eq.\ (\ref{eq:shearflow}). However, now the shear flow is no more constrained and will self-consistently dissipate with time. For simplicity, we still apply periodic boundary conditions, which means the $\alpha$-particles generated by fusion reactions will anyhow deposit in the hot spot. This can overestimate the fusion yield since in reality the 3.5 MeV $\alpha$-particles generated by DT reactions have a very large $\lambda$, and part of them will inevitably fly out of the hot spot. As for simulation parameters, we still adopt the parameters used in Sec.\ \ref{sec:tail} to facilitate comparison. That is, a fiducial case with $T=4$ keV, equimolar D-T mixture with $\rho=150$ g/cc, and a shear flow with $u_0=\sqrt{3/2}v_{th}$ and $L=0.56\ \mu\text{m}$. Supplemental simulations with $T=4$ keV and 5 keV but no shear flow are also conducted as benchmarks. Another simulation with $L=1.12\ \mu\text{m}$ is conducted for comparison. The two TKE simulations have the same initial total energy as the 5 keV benchmark. In each simulation, we use a spatial grid $\Delta z=0.002\ \mu\rm m$, and assign 1000 macro-particles per cell to sample each species. Bremsstrahlung and other types of radiations are not contained in our simulation.

\subsection{Evolution of the flow and the tail distribution}

\begin{figure}[b]
\includegraphics[width=0.98 \columnwidth]{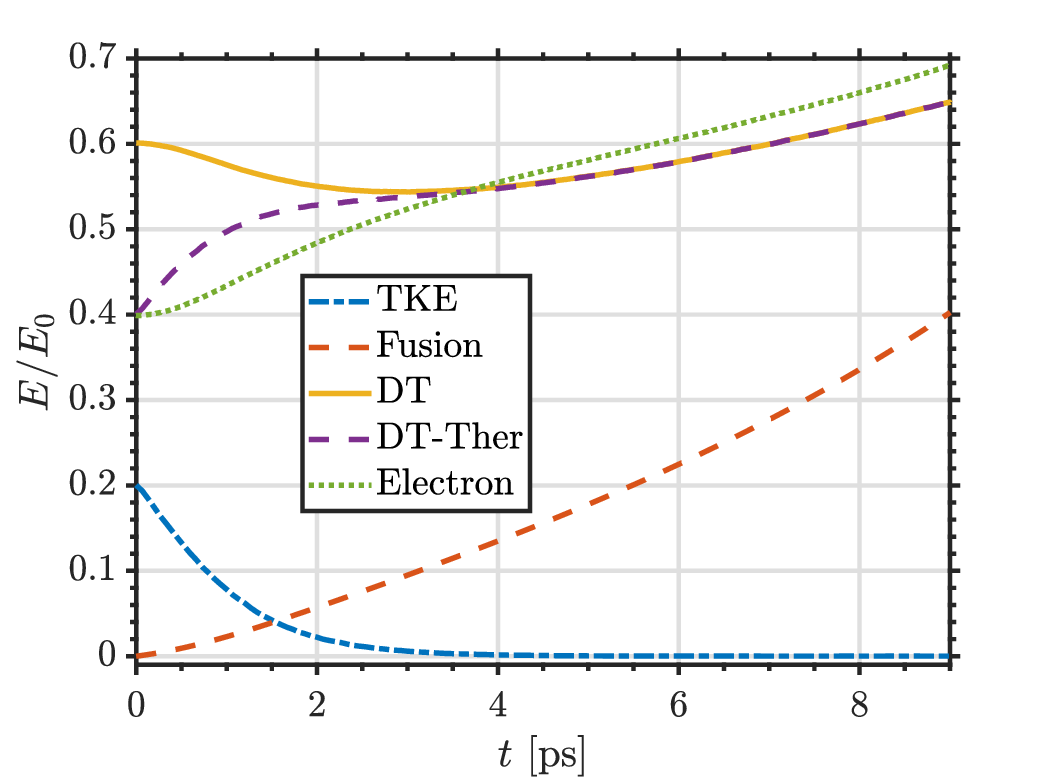}
\caption{\label{fig:energy}Evolution of different forms of energies in the fiducial case. Displayed are TKE (blue), cumulative fusion-released energy carried by $\alpha$-particles (red), total energy of D\&T ions (yellow), thermal energies of D\&T ions (purple, which equals total energy of D\&T ions minus TKE) and total energy of electrons (purple). All energies are normalized by the initial total energy of the system.}
\end{figure}

Now we display results from the PIC simulations. Figure\ \ref{fig:energy} shows the evolution of different forms of energies in the fiducial PIC simulation. The TKE in the system is evaluated as $\text{TKE}=\int dV(\rho_D u_D^2+\rho_T u_T^2)/2$. Initially, the flow amplitude is $u_0 = \sqrt{3/2}v_{th}$, corresponding to a TKE which constitutes $20\%$ of the total system energy. As the simulation begins, TKE starts to dissipate rapidly. Consistent with the viscous dissipation estimate in Sec.\ \ref{sec:tail} ($\tau_v=1/\eta k^2\approx1.8$ ps), the TKE decays to about $15\%$ of its initial value within the first $1.8$ ps and becomes fully exhausted by $t\approx 4$ ps. Consequently, the heating process of the system clearly exhibits a two-stage dynamics. In the first viscous-dominated phase ($t\lesssim 2.5\ \rm ps$), since $\nu_{ii}\gg\nu_{ie}$, TKE is rapidly converted into thermal energy of the DT ions, which reaches 0.5$E_0$ at about $t=1\ \rm ps$. This corresponds to an effective temperature $T_{\rm eff}=5$\ keV of the DT ions. Meanwhile, in this process electrons are also slowly heated through collisions with ions, leading to a decrease in the total energy of DT ions. After the TKE vanishes, heating by fusion generated $\alpha$-particles features the second stage ($t \gtrsim 2.5$ ps). Since $v_{thi}\ll v_{the}\sim v_{\alpha}$, electrons have a larger frequency with the 3.5 MeV $\alpha$-particles. In consequence, in this stage $\alpha$-particles mainly transfer their energy to electrons, while DT ions are only heated by the medium of electrons and slowed-down $\alpha$-particles. This leads to a much slower heating of DT ions compared to in the first stage. Also, the energy and effective temperature of electrons overtake those of DT ions in this stage.

\begin{figure}[b]
\includegraphics[width=0.98 \columnwidth]{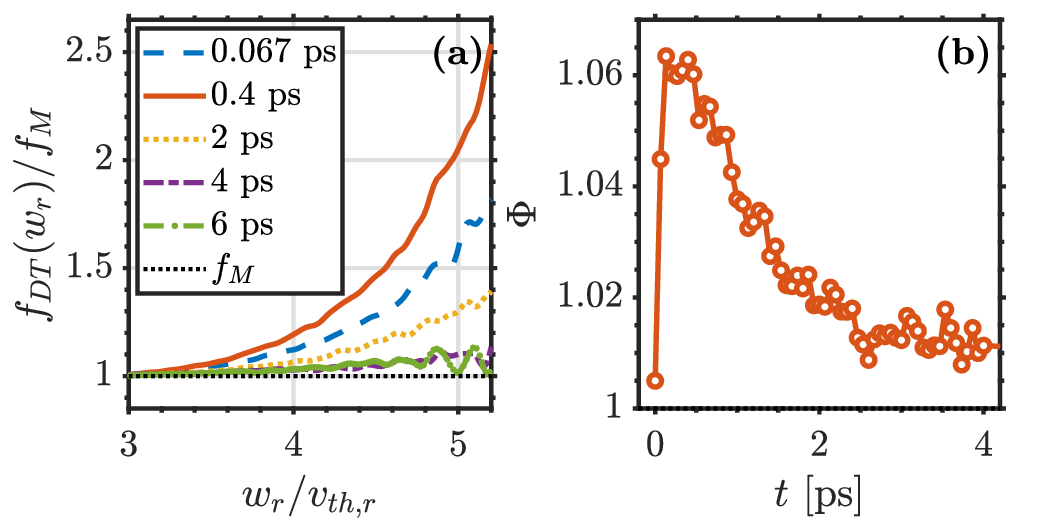}
\caption{\label{fig:pictail}(a) The spatial averaged tail distributions of the relative velocity, normalized by their Maxwellian counterpart, $f(w_r)/f_M$, extracted at 0.067 ps (blue), 0.4 ps (red), 2 ps (yellow), 4 ps (purple), and 6 ps (green) respectively. (b) The corresponding SFRE $\Phi=\Sigma[f,f]/\Sigma[f_M,f_M]$ evaluated at each time instant.}
\end{figure}

Figure\ \ref{fig:energy} alone does not demonstrate SFRE. To illustrate the tail enhancement, we track the information of all particles in the simulation at an interval of $\Delta t=0.067\ \rm ps$, and calculate a distribution function of the relative velocity $f(w_r)$ using Eqs.\ (\ref{eq:conv}) and (\ref{eq:wr}). Figure \ref{fig:pictail}(a) displays the spatially averaged relative velocity distribution $f(w_r)$ at five characteristic instants, each normalized by a Maxwellian with the same energy. Corresponding $\Phi=\Sigma[f,f]/\Sigma[f_M,f_M]$ evaluated at each time instant is also displayed in Panel (b). At a very early time $t=0.067$ ps (see the blue dashed line), a non-Maxwellian tail already emerges, although this time is not even long enough for tail ions to cross a quarter of the wavelength. The most pronounced non-Maxwellian tail appears at $t = 0.4$ ps, and yields a maximum $\Phi\approx1.06$ . This timescale is comparable to the transit time of tail ions across the wavelength, $\tau_{\rm t} \sim L/v_p \approx 0.34$ ps, indicating that the tail distribution is  established as soon as fast ions complete their first few traversals from the wave crest to the trough. 

After $t \approx 0.4$ ps, $f(w_r)/f_M$ begins to decrease rapidly. However, we note that this decrease is partly due to the increasing temperature of the DT ions. Remember that as TKE dissipates quickly,  the effective temperature of DT ions increase to 5 keV at about $t=1$ ps. This leads to a $f_M$ with a larger temperature and weakens the relative amplitude of the tail enhancement in $f(w_r)/f_M$, while the absolute value of the tail distortion may still increase. Also, this results in a maximum $\Phi=1.06$, which is much smaller than our prediction in Sec.\ \ref{sec:tail}. However, at $t=1$ ps where $T_{\rm eff}\approx5$ keV, $\Phi$ still maintains a value at about 1.04, which is not negligible. 

At $t=2$ ps, which is close to the shear flow dissipation timescale $\tau_v=1.8$ ps, the tail still retains an relative amplitude to about a half of the highest tail at $t=0.4\ \rm ps$. This also results from the low collision frequency of the tail particles. Since $\nu \propto v^{-3}$, the relaxation time of the tail is roughly $\tau_{tail} \sim (v_p/v_{th})^3 \tau_v$, which makes them respond to the dissipated flow slower and maintain the tail distribution for a certain timescale after the shear flow has dissipated. After that, the tail distribution continues to diminish. At $t = 6$ ps, the distribution has mostly relaxed into a Maxwellian, although some slight distortions still occur on the tail, partly because of the finite numbers of the macro-particles in the simulation.
Compared with the steady-state FP result in Fig.\ \ref{fig:1ddis}(c), the PIC tail enhancement is not only weaker in relative amplitude, but also shifted to larger $w_r/v_{th}$. Besides the mentioned heating of the DT ions, the inclusion of the third velocity dimension $v_y$ in the PIC simulation also contributes to the difference. The additional degree of freedom recovers the full 3V pitch-angle scattering that is absent in the 1D2V FP solver, providing a more efficient isotropization channel. Furthermore, the convolution integral in Eq.\ (\ref{eq:conv}) operates in a higher-dimensional phase space, which topologically dilutes the relative-velocity tail.

\subsection{Evolution of fusion reactivity}

\begin{figure}[b]
\includegraphics[width=0.98 \columnwidth]{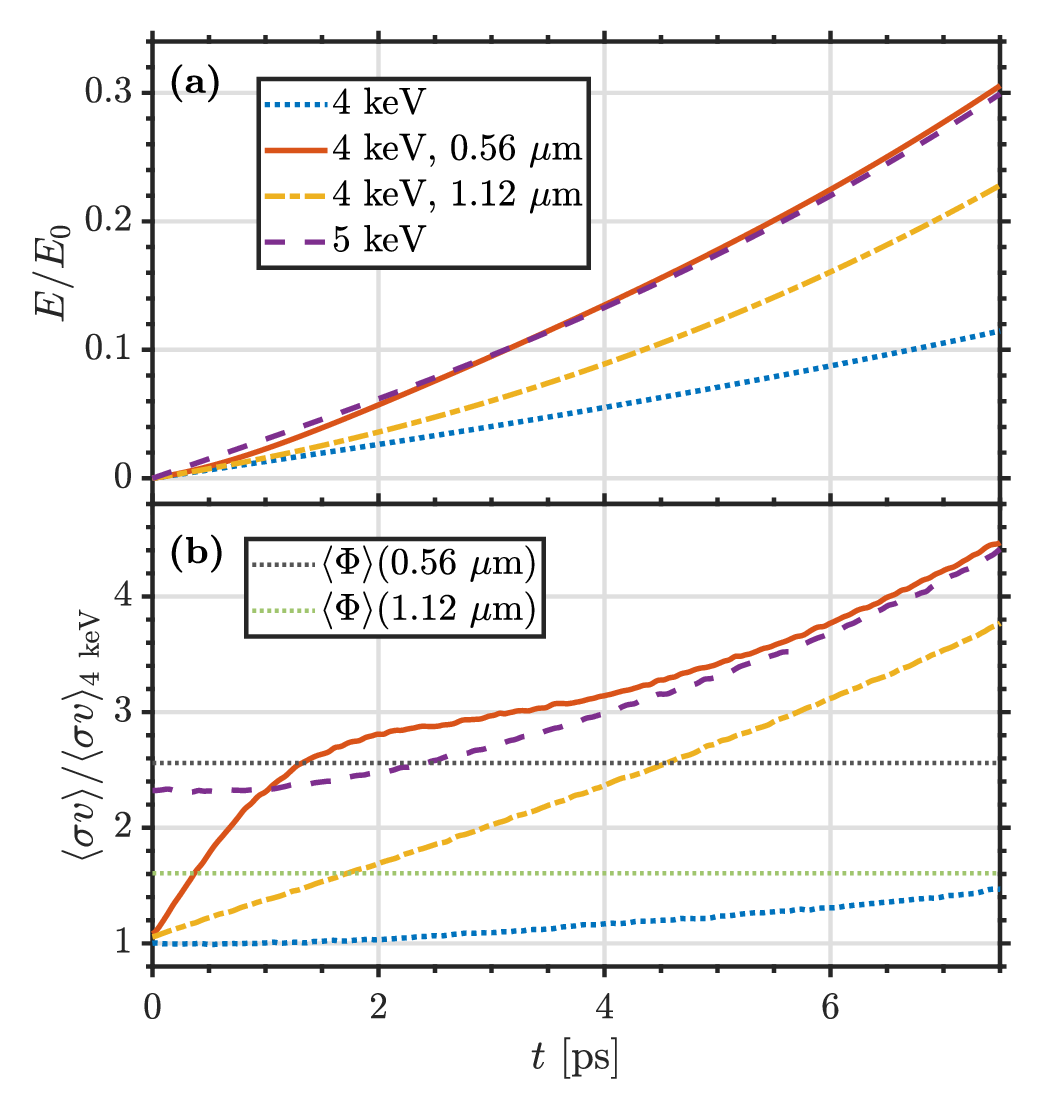}
\caption{\label{fig:yield}(a) Cumulative fusion-energy carried by $\alpha$-particles in different simulations, normalized by the initial total energy of the 5 keV benchmark. Displayed are simulations of 4 keV and no TKE (blue), 4 keV and TKE with $L=0.56\ \mu\text {m}$ (red), 4 keV and TKE with $L=1.12\ \mu\text {m}$ (yellow), and 5 keV and no TKE (purple). (b) Corresponding fusion reactivities, normalized by the 4 keV reactivity. $\langle\Phi\rangle$ calculated for the two TKE cases using the method in Sec.\ \ref{sec:tail} are also displayed.}
\end{figure}

The cumulative energy released by fusion reactions in the simulation is displayed in Fig.\ \ref{fig:yield}(a), where the other three simulations are also shown for comparison. In Panel (b) we show the corresponding fusion reactivity $\langle \sigma v\rangle$, normalized by the initial 4 keV reactivity. This value is equivalent to a time-dependent $\langle \Phi\rangle$.  $\langle \Phi\rangle$ calculated in Sec.\ \ref{sec:tail} for the 2 TKE cases are also displayed for comparison. 

In the 2 TKE cases, $\langle\Phi\rangle$ naturally starts from 1 because the tail distribution has not formed yet at the beginning. After that, $\langle\Phi\rangle$ quickly increases due to the formation of the tail distribution and also dissipation of the TKE. In consequence, both simulations with TKE consistently outperform the 4 keV Maxwellian benchmark throughout the entire simulation. For the same reason, the $L=0.56\ \mu \rm m$ case yields a larger reactivity than the $L=1.12\ \mu\rm m$ case, due to a larger tail enhancement and a faster dissipation of TKE. The larger reactivity results in generation of more $\alpha$-particles and heating of the system, which in turn increases the reactivity and forms a positive feedback loop. In consequence, among these 3 simulations the gap in fusion yield continues to widen throughout the simulation.

However, a striking exception occurs between the case of 4 keV, 0.56 $\mu$m and the 5 keV benchmark. Although the 5 keV benchmark yields a large $\Phi=2.32$ at the beginning, the reactivity in the 4 keV, 0.56 $\mu$m case quickly increases and overtakes the 5 keV benchmark at about 1.05 ps, before the complete dissipation of the shear flow. This overtake is consistent with our prediction in Sec.\ \ref{sec:pic} that $\langle\Phi\rangle(0.56\ \mu\mathrm{m})\approx2.56>\Phi_{th}=2.31$. At around 1.8 ps, the difference in the reactivity is even larger than our prediction. The reason is two-fold: on one hand, TKE dissipates more into thermal energy of ions than that of electrons, generating a DT plasma at around $T_{\rm eff}\approx 5.2\ \rm keV$ (see Fig.\ \ref{fig:energy}). Meanwhile, in the 5 keV benchmark only a little part of the energy released by fusion has converted into the thermal energy of DT ions, where $T_{\rm eff}\approx 5.1\ \rm keV$.  On the other hand, while TKE dissipates into thermal energy and increases temperature of the system, the tail distribution formed earlier still remains, maintaining a $\Phi \approx1.02$ (see Fig.\ \ref{fig:pictail}). Both effects combined together result in an enhancement even larger than the steady-state prediction.

This anomalously large enhancement also indicates that, in spite of the kinetic aspect of the SFRE, partitioning energy into TKE can also  be thermodynamically beneficial since TKE is preferentially dissipated into thermal energy of ions. This preferential heating decouples  $T_i$ from $T_e$. While a higher $T_i$ provides a larger reactivity, a lower $T_e$ also reduces the loss of bremsstrahlung, which is both advantageous for fusion design.

After 1.8 ps, the TKE case keeps a larger reactivity and achieves a comeback in the total fusion yield at about 3.24 ps. Since then, the positive feedback loop continues to act and maintains the lead of the TKE case. Although the difference in the final yield is smaller than the steady-state prediction, this result is a striking and promising evidence of SFRE, where between the two cases with the same initial total energy, the case with a lower initial temperature yields a larger fusion energy in total.

In contrast, the case with $1.12\ \mu\rm m$ never catches up with the 5 keV benchmark, which is also consistent with our prediction in Sec.\ \ref{sec:tail} that $\langle\Phi\rangle(1.12\ \mu\mathrm{m})\approx1.61<\Phi_{th}=2.31$. The relatively slow dissipation of TKE can not reverse the difference in this case. After a long enough timescale, the shear flow and the tail distortion will disappear in both TKE cases, and they will  perform as thermalized burning hot spots.

Taken together, the PIC results paint a self-consistent picture of SFRE in a dynamically evolving hot spot. While the enhancement to the tail distribution is more modest than the steady-state BGK predictions, the distorted tail can retain a longer timescale even after the shear flow dissipates. Also, TKE is preferentially dissipated into thermal energy of ions, increasing the effective temperature of ions substantially. As a result, the transient fusion reactivity can even be larger than the steady-state prediction. If the enhancement is large enough, the reactivity and resultant fusion yield can indeed exceed those of the thermalized counterpart. That is to say, in principle, prescribing turbulence or shear flow in the hot spot can be more efficient than heating it with the same energy.

\section{Conclusion}\label{sec:con}
Following the work of Fetsch and Fisch \cite{fetsch2025enhancement,fetsch2025analytical}, in this paper, we investigated the shear flow reactivity enhancement effect with a series of numerical simulations. We demonstrate that the modified BGK operator significantly overestimates the SFRE, whereas the more physically accurate FP operator and integrated PIC simulations predict only a modest enhancement. In PIC simulations, we also observe a combined effect of shear flow dissipation and tail enhancement, which even enlarges SFRE compared to the steady-state prediction.

With a simplified 1D2V kinetic solver, the steady-state ion tail distributions driven by a sinusoidal shear flow were solved using both the BGK and FP operators. We find that the FP operator consistently predicts a more isotropic and less distorted distribution in the velocity phase space, which yields a smaller tail distribution of the relative velocity between reactants, and henceforth a smaller fusion reactivity compared to the BGK model. For typical ICF parameters, $\rho=150$ g/cc, $T=4$ keV and assuming a shear flow with the wavelength $L=0.56\ \rm \mu m$, the FP operator predicts a SFRE $\langle\Phi\rangle=2.56$ slightly larger than the thermalized counterpart $\Phi_{th}=2.31$, while the BGK operator overestimates $\langle\Phi\rangle$ up to 4.58. Quite generally, SFRE is a monotonic function of both the collision frequency $\nu_0$ and the flow amplitude $u_0/v_{th}$, while it is a non-monotonic function of the ion temperature $T$. This non-monotonicity suggests the existence of an optimal temperature for maximizing the SFRE, which requires shear flow or turbulence in hot spots to be sustained over a sufficient timescale or generated at a relatively late stage. Interestingly, as the collision frequency increases, we observe a spatial phase-reversal in the reactivity enhancement factor $\Phi(z)$, which results from the transition of the tail particle transport from global to near-local regimes. 

We also conducted 1D3V PIC simulations incorporating fusion reactions to evaluate the significance of SFRE in a time-dependent burning process. In the simulation with the same parameters adopted in the steady-state solver, $\rho=150$ g/cc, $T=4$ keV and $L=0.56\ \rm \mu m$, we observe a tail distortion with a much smaller relative amplitude than the steady-state prediction. This stems from two aspects: first, in PIC simulations the shear flow dissipates rapidly, increasing the temperature of the system. This weakens the relative amplitude of the distorted tail distribution. Second, in the PIC simulation the full 3 dimensions of the velocity space are included, which correctly recovers the pitch-angle scattering underestimated in the 1D2V steady-state solver.  However, we also observe that the enhanced tail distribution can maintain a longer timescale than the dissipation timescale of the shear flow. Meanwhile, the TKE of the shear flow is preferentially dissipated into the thermal energy of ions, which means in a certain time duration the effective temperature of the ions can exceed that in the case where all energy is distributed equally into thermal energy of ions and electrons. Combined together, this yields a SFRE $\Phi$ in the PIC simulation even larger than the steady-state prediction. In consequence, the reactivity in the 4 keV TKE simulation overtakes that in the 5 keV benchmark, and eventually yields a larger fusion energy. Although the model of the simulation is idealistic, this result is a direct and encouraging evidence of SFRE, suggesting that partitioning energy into TKE in the hot spot can indeed increase fusion reactivity.

We would like to note again that, there are several limitations of our work. First, our results in Sec.\ \ref{sec:tail} were derived in a 2V space. This does not affect the BGK results because the uninvolved velocity $v_y$ is strictly retained with the BGK operator. However, for the FP operator, neglecting $v_y$ means the pitch-angle scattering is restricted, thus overestimating the extent of distortion of the tail distribution. Therefore, Fig.\ \ref{fig:scan2d} should be considered as an upper limit for the actual SFRE $\Phi$. Second, we only considered a 1D space along the shear flow wavevector. In reality the turbulent flow can be much more complicated topologically, and distributing TKE into all dimensions will likely reduce the SFRE. Third, in our paper we consistently assumed a single-mode shear flow, while the real turbulence may take a Kolmogorov-like spectrum. Also, in real turbulence the energy is continuously cascaded from integral scales down to dissipation scales. Resolving the second and third limitations demands a more detailed description of the turbulent flow, which depends on analyses on specific problems. Fourth, periodic boundary conditions are applied to both the steady-state solver in Sec.\ \ref{sec:tail} and the PIC simulations in Sec.\ \ref{sec:pic}. This avoids the problem of maintaining the shear flow and confining the $\alpha$-particles, which can lead to overestimate of the fusion production. However, this does not affect the physical essence of SFRE explored in this work.

In this paper we stressed the comparison between the SFRE $\Phi$ and its thermalized counterpart $\Phi_{th}$. However, $\Phi$ does not need to be literally larger than $\Phi_{th}$ to create benefits. As stated by FF, partitioning energy into TKE rather than full thermalization reduces the bremsstrahlung and also the stopping distance of $\alpha$-particles, which may allow a smaller fuel target, a smaller driving energy and is advantageous for practical fusion design. In consequence, it is a comprehensive question whether a fusion scheme with TKE intentionally driven can be more effective in fusion power generation. Up to date, there is no easy answer to this question since no one had designed a turbulent hot spot intentionally. However, at least we can conclude that TKE inherent in fusion schemes nowadays is not totally harmful.

\begin{acknowledgments}
This work was supported by Science Challenge Project (No.\ TZ2025012), the Strategic Priority Research Program of Chinese Academy of Sciences (Grant No.\ XDA25010100 and XDA25050500),  National Natural Science Foundation of China (Grant No.\ 12075204) and Shanghai Municipal Science and Technology Key Project (Grant No.\ 22JC1401500). Dong Wu thanks the sponsorship from Yangyang Development Fund.
\end{acknowledgments}



\bibliography{apssamp}

\providecommand{\noopsort}[1]{}\providecommand{\singleletter}[1]{#1}%
\begin{thebibliography}{43}%
\makeatletter
\providecommand \@ifxundefined [1]{%
 \@ifx{#1\undefined}
}%
\providecommand \@ifnum [1]{%
 \ifnum #1\expandafter \@firstoftwo
 \else \expandafter \@secondoftwo
 \fi
}%
\providecommand \@ifx [1]{%
 \ifx #1\expandafter \@firstoftwo
 \else \expandafter \@secondoftwo
 \fi
}%
\providecommand \natexlab [1]{#1}%
\providecommand \enquote  [1]{``#1''}%
\providecommand \bibnamefont  [1]{#1}%
\providecommand \bibfnamefont [1]{#1}%
\providecommand \citenamefont [1]{#1}%
\providecommand \href@noop [0]{\@secondoftwo}%
\providecommand \href [0]{\begingroup \@sanitize@url \@href}%
\providecommand \@href[1]{\@@startlink{#1}\@@href}%
\providecommand \@@href[1]{\endgroup#1\@@endlink}%
\providecommand \@sanitize@url [0]{\catcode `\\12\catcode `\$12\catcode
  `\&12\catcode `\#12\catcode `\^12\catcode `\_12\catcode `\%12\relax}%
\providecommand \@@startlink[1]{}%
\providecommand \@@endlink[0]{}%
\providecommand \url  [0]{\begingroup\@sanitize@url \@url }%
\providecommand \@url [1]{\endgroup\@href {#1}{\urlprefix }}%
\providecommand \urlprefix  [0]{URL }%
\providecommand \Eprint [0]{\href }%
\providecommand \doibase [0]{https://doi.org/}%
\providecommand \selectlanguage [0]{\@gobble}%
\providecommand \bibinfo  [0]{\@secondoftwo}%
\providecommand \bibfield  [0]{\@secondoftwo}%
\providecommand \translation [1]{[#1]}%
\providecommand \BibitemOpen [0]{}%
\providecommand \bibitemStop [0]{}%
\providecommand \bibitemNoStop [0]{.\EOS\space}%
\providecommand \EOS [0]{\spacefactor3000\relax}%
\providecommand \BibitemShut  [1]{\csname bibitem#1\endcsname}%
\let\auto@bib@innerbib\@empty
\bibitem [{\citenamefont {Nuckolls}\ \emph {et~al.}(1972)\citenamefont
  {Nuckolls}, \citenamefont {Wood}, \citenamefont {Thiessen},\ and\
  \citenamefont {Zimmerman}}]{nuckolls1972laser}%
  \BibitemOpen
  \bibfield  {author} {\bibinfo {author} {\bibfnamefont {J.}~\bibnamefont
  {Nuckolls}}, \bibinfo {author} {\bibfnamefont {L.}~\bibnamefont {Wood}},
  \bibinfo {author} {\bibfnamefont {A.}~\bibnamefont {Thiessen}},\ and\
  \bibinfo {author} {\bibfnamefont {G.}~\bibnamefont {Zimmerman}},\ }\bibfield
  {title} {\bibinfo {title} {Laser compression of matter to super-high
  densities: Thermonuclear (ctr) applications},\ }\href
  {https://doi.org/10.1038/239139a0} {\bibfield  {journal} {\bibinfo  {journal}
  {Nature}\ }\textbf {\bibinfo {volume} {239}},\ \bibinfo {pages} {139}
  (\bibinfo {year} {1972})}\BibitemShut {NoStop}%
\bibitem [{\citenamefont {Lindl}\ \emph {et~al.}(2004)\citenamefont {Lindl},
  \citenamefont {Amendt}, \citenamefont {Berger}, \citenamefont {Glendinning},
  \citenamefont {Glenzer}, \citenamefont {Haan}, \citenamefont {Kauffman},
  \citenamefont {Landen},\ and\ \citenamefont {Suter}}]{Lindl2004}%
  \BibitemOpen
  \bibfield  {author} {\bibinfo {author} {\bibfnamefont {J.~D.}\ \bibnamefont
  {Lindl}}, \bibinfo {author} {\bibfnamefont {P.}~\bibnamefont {Amendt}},
  \bibinfo {author} {\bibfnamefont {R.~L.}\ \bibnamefont {Berger}}, \bibinfo
  {author} {\bibfnamefont {S.~G.}\ \bibnamefont {Glendinning}}, \bibinfo
  {author} {\bibfnamefont {S.~H.}\ \bibnamefont {Glenzer}}, \bibinfo {author}
  {\bibfnamefont {S.~W.}\ \bibnamefont {Haan}}, \bibinfo {author}
  {\bibfnamefont {R.~L.}\ \bibnamefont {Kauffman}}, \bibinfo {author}
  {\bibfnamefont {O.~L.}\ \bibnamefont {Landen}},\ and\ \bibinfo {author}
  {\bibfnamefont {L.~J.}\ \bibnamefont {Suter}},\ }\bibfield  {title} {\bibinfo
  {title} {The physics basis for ignition using indirect-drive targets on the
  national ignition facility},\ }\href {https://doi.org/10.1063/1.1578638}
  {\bibfield  {journal} {\bibinfo  {journal} {Physics of Plasmas}\ }\textbf
  {\bibinfo {volume} {11}},\ \bibinfo {pages} {339} (\bibinfo {year}
  {2004})}\BibitemShut {NoStop}%
\bibitem [{\citenamefont {Betti}\ and\ \citenamefont
  {Hurricane}(2016)}]{betti2016inertial}%
  \BibitemOpen
  \bibfield  {author} {\bibinfo {author} {\bibfnamefont {R.}~\bibnamefont
  {Betti}}\ and\ \bibinfo {author} {\bibfnamefont {O.}~\bibnamefont
  {Hurricane}},\ }\bibfield  {title} {\bibinfo {title} {Inertial-confinement
  fusion with lasers},\ }\href {https://doi.org/10.1038/nphys3736} {\bibfield
  {journal} {\bibinfo  {journal} {Nature Physics}\ }\textbf {\bibinfo {volume}
  {12}},\ \bibinfo {pages} {435} (\bibinfo {year} {2016})}\BibitemShut
  {NoStop}%
\bibitem [{\citenamefont {Marinak}\ \emph {et~al.}(2001)\citenamefont
  {Marinak}, \citenamefont {Kerbel}, \citenamefont {Gentile}, \citenamefont
  {Jones}, \citenamefont {Munro}, \citenamefont {Pollaine}, \citenamefont
  {Dittrich},\ and\ \citenamefont {Haan}}]{Marinak2001}%
  \BibitemOpen
  \bibfield  {author} {\bibinfo {author} {\bibfnamefont {M.~M.}\ \bibnamefont
  {Marinak}}, \bibinfo {author} {\bibfnamefont {G.~D.}\ \bibnamefont {Kerbel}},
  \bibinfo {author} {\bibfnamefont {N.~A.}\ \bibnamefont {Gentile}}, \bibinfo
  {author} {\bibfnamefont {O.}~\bibnamefont {Jones}}, \bibinfo {author}
  {\bibfnamefont {D.}~\bibnamefont {Munro}}, \bibinfo {author} {\bibfnamefont
  {S.}~\bibnamefont {Pollaine}}, \bibinfo {author} {\bibfnamefont {T.~R.}\
  \bibnamefont {Dittrich}},\ and\ \bibinfo {author} {\bibfnamefont {S.~W.}\
  \bibnamefont {Haan}},\ }\bibfield  {title} {\bibinfo {title}
  {Three-dimensional hydra simulations of national ignition facility targets},\
  }\href {https://doi.org/10.1063/1.1356740} {\bibfield  {journal} {\bibinfo
  {journal} {Physics of Plasmas}\ }\textbf {\bibinfo {volume} {8}},\ \bibinfo
  {pages} {2275} (\bibinfo {year} {2001})}\BibitemShut {NoStop}%
\bibitem [{\citenamefont {Clark}\ \emph {et~al.}(2019)\citenamefont {Clark},
  \citenamefont {Weber}, \citenamefont {Milovich}, \citenamefont {Pak},
  \citenamefont {Casey}, \citenamefont {Hammel} \emph {et~al.}}]{Clark2019}%
  \BibitemOpen
  \bibfield  {author} {\bibinfo {author} {\bibfnamefont {D.~S.}\ \bibnamefont
  {Clark}}, \bibinfo {author} {\bibfnamefont {C.~R.}\ \bibnamefont {Weber}},
  \bibinfo {author} {\bibfnamefont {J.~L.}\ \bibnamefont {Milovich}}, \bibinfo
  {author} {\bibfnamefont {A.~E.}\ \bibnamefont {Pak}}, \bibinfo {author}
  {\bibfnamefont {D.~T.}\ \bibnamefont {Casey}}, \bibinfo {author}
  {\bibfnamefont {B.~A.}\ \bibnamefont {Hammel}}, \emph {et~al.},\ }\bibfield
  {title} {\bibinfo {title} {Three-dimensional modeling and hydrodynamic
  scaling of national ignition facility implosions},\ }\href
  {https://doi.org/10.1063/1.5091449} {\bibfield  {journal} {\bibinfo
  {journal} {Physics of Plasmas}\ }\textbf {\bibinfo {volume} {26}},\ \bibinfo
  {pages} {050601} (\bibinfo {year} {2019})}\BibitemShut {NoStop}%
\bibitem [{\citenamefont {Spitzer}(2006)}]{spitzer2006physics}%
  \BibitemOpen
  \bibfield  {author} {\bibinfo {author} {\bibfnamefont {L.}~\bibnamefont
  {Spitzer}},\ }\href@noop {} {\emph {\bibinfo {title} {Physics of fully
  ionized gases}}}\ (\bibinfo  {publisher} {Courier Corporation},\ \bibinfo
  {year} {2006})\BibitemShut {NoStop}%
\bibitem [{\citenamefont {Kritcher}\ \emph {et~al.}(2014)\citenamefont
  {Kritcher}, \citenamefont {Town}, \citenamefont {Bradley}, \citenamefont
  {Clark}, \citenamefont {Spears}, \citenamefont {Jones}, \citenamefont {Haan},
  \citenamefont {Springer}, \citenamefont {Lindl}, \citenamefont {Scott},
  \citenamefont {Callahan}, \citenamefont {Edwards},\ and\ \citenamefont
  {Landen}}]{Kritcher2014}%
  \BibitemOpen
  \bibfield  {author} {\bibinfo {author} {\bibfnamefont {A.~L.}\ \bibnamefont
  {Kritcher}}, \bibinfo {author} {\bibfnamefont {R.}~\bibnamefont {Town}},
  \bibinfo {author} {\bibfnamefont {D.}~\bibnamefont {Bradley}}, \bibinfo
  {author} {\bibfnamefont {D.}~\bibnamefont {Clark}}, \bibinfo {author}
  {\bibfnamefont {B.}~\bibnamefont {Spears}}, \bibinfo {author} {\bibfnamefont
  {O.}~\bibnamefont {Jones}}, \bibinfo {author} {\bibfnamefont
  {S.}~\bibnamefont {Haan}}, \bibinfo {author} {\bibfnamefont {P.~T.}\
  \bibnamefont {Springer}}, \bibinfo {author} {\bibfnamefont {J.}~\bibnamefont
  {Lindl}}, \bibinfo {author} {\bibfnamefont {R.~H.~H.}\ \bibnamefont {Scott}},
  \bibinfo {author} {\bibfnamefont {D.}~\bibnamefont {Callahan}}, \bibinfo
  {author} {\bibfnamefont {M.~J.}\ \bibnamefont {Edwards}},\ and\ \bibinfo
  {author} {\bibfnamefont {O.~L.}\ \bibnamefont {Landen}},\ }\bibfield  {title}
  {\bibinfo {title} {Metrics for long wavelength asymmetries in inertial
  confinement fusion implosions on the national ignition facility},\ }\href
  {https://doi.org/10.1063/1.4871718} {\bibfield  {journal} {\bibinfo
  {journal} {Physics of Plasmas}\ }\textbf {\bibinfo {volume} {21}},\ \bibinfo
  {pages} {042708} (\bibinfo {year} {2014})}\BibitemShut {NoStop}%
\bibitem [{\citenamefont {Zhou}(2017{\natexlab{a}})}]{ZHOU20170}%
  \BibitemOpen
  \bibfield  {author} {\bibinfo {author} {\bibfnamefont {Y.}~\bibnamefont
  {Zhou}},\ }\bibfield  {title} {\bibinfo {title} {Rayleigh–taylor and
  richtmyer–meshkov instability induced flow, turbulence, and mixing. i},\
  }\href {https://doi.org/https://doi.org/10.1016/j.physrep.2017.07.005}
  {\bibfield  {journal} {\bibinfo  {journal} {Physics Reports}\ }\textbf
  {\bibinfo {volume} {720-722}},\ \bibinfo {pages} {1} (\bibinfo {year}
  {2017}{\natexlab{a}})},\ \bibinfo {note} {rayleigh-Taylor and
  Richtmyer-Meshkov instability induced flow, turbulence, and mixing.
  I}\BibitemShut {NoStop}%
\bibitem [{\citenamefont {Zhou}(2017{\natexlab{b}})}]{ZHOU20171}%
  \BibitemOpen
  \bibfield  {author} {\bibinfo {author} {\bibfnamefont {Y.}~\bibnamefont
  {Zhou}},\ }\bibfield  {title} {\bibinfo {title} {Rayleigh–taylor and
  richtmyer–meshkov instability induced flow, turbulence, and mixing. ii},\
  }\href {https://doi.org/https://doi.org/10.1016/j.physrep.2017.07.008}
  {\bibfield  {journal} {\bibinfo  {journal} {Physics Reports}\ }\textbf
  {\bibinfo {volume} {723-725}},\ \bibinfo {pages} {1} (\bibinfo {year}
  {2017}{\natexlab{b}})},\ \bibinfo {note} {rayleigh–Taylor and
  Richtmyer–Meshkov instability induced flow, turbulence, and mixing.
  II}\BibitemShut {NoStop}%
\bibitem [{\citenamefont {Guo}\ \emph {et~al.}(2024)\citenamefont {Guo},
  \citenamefont {Wu},\ and\ \citenamefont {Zhang}}]{Guo2024}%
  \BibitemOpen
  \bibfield  {author} {\bibinfo {author} {\bibfnamefont {Y.}~\bibnamefont
  {Guo}}, \bibinfo {author} {\bibfnamefont {D.}~\bibnamefont {Wu}},\ and\
  \bibinfo {author} {\bibfnamefont {J.}~\bibnamefont {Zhang}},\ }\bibfield
  {title} {\bibinfo {title} {Effects of mass diffusion on rayleigh–taylor
  instability under a large gravity},\ }\href
  {https://doi.org/10.1063/5.0234173} {\bibfield  {journal} {\bibinfo
  {journal} {Physics of Plasmas}\ }\textbf {\bibinfo {volume} {31}},\ \bibinfo
  {pages} {112106} (\bibinfo {year} {2024})}\BibitemShut {NoStop}%
\bibitem [{\citenamefont {Springer}\ \emph {et~al.}(2018)\citenamefont
  {Springer}, \citenamefont {Hurricane}, \citenamefont {Hammer}, \citenamefont
  {Betti}, \citenamefont {Callahan} \emph {et~al.}}]{Springer_2019}%
  \BibitemOpen
  \bibfield  {author} {\bibinfo {author} {\bibfnamefont {P.}~\bibnamefont
  {Springer}}, \bibinfo {author} {\bibfnamefont {O.}~\bibnamefont {Hurricane}},
  \bibinfo {author} {\bibfnamefont {J.}~\bibnamefont {Hammer}}, \bibinfo
  {author} {\bibfnamefont {R.}~\bibnamefont {Betti}}, \bibinfo {author}
  {\bibfnamefont {D.}~\bibnamefont {Callahan}}, \emph {et~al.},\ }\bibfield
  {title} {\bibinfo {title} {A 3d dynamic model to assess the impacts of
  low-mode asymmetry, aneurysms and mix-induced radiative loss on capsule
  performance across inertial confinement fusion platforms},\ }\href
  {https://doi.org/10.1088/1741-4326/aaed65} {\bibfield  {journal} {\bibinfo
  {journal} {Nuclear Fusion}\ }\textbf {\bibinfo {volume} {59}},\ \bibinfo
  {pages} {032009} (\bibinfo {year} {2018})}\BibitemShut {NoStop}%
\bibitem [{\citenamefont {Kline}\ \emph {et~al.}(2019)\citenamefont {Kline},
  \citenamefont {Batha}, \citenamefont {Benedetti}, \citenamefont {Bennett},
  \citenamefont {Bhandarkar} \emph {et~al.}}]{Kline_2019}%
  \BibitemOpen
  \bibfield  {author} {\bibinfo {author} {\bibfnamefont {J.}~\bibnamefont
  {Kline}}, \bibinfo {author} {\bibfnamefont {S.}~\bibnamefont {Batha}},
  \bibinfo {author} {\bibfnamefont {L.}~\bibnamefont {Benedetti}}, \bibinfo
  {author} {\bibfnamefont {D.}~\bibnamefont {Bennett}}, \bibinfo {author}
  {\bibfnamefont {S.}~\bibnamefont {Bhandarkar}}, \emph {et~al.},\ }\bibfield
  {title} {\bibinfo {title} {Progress of indirect drive inertial confinement
  fusion in the united states},\ }\href
  {https://doi.org/10.1088/1741-4326/ab1ecf} {\bibfield  {journal} {\bibinfo
  {journal} {Nuclear Fusion}\ }\textbf {\bibinfo {volume} {59}},\ \bibinfo
  {pages} {112018} (\bibinfo {year} {2019})}\BibitemShut {NoStop}%
\bibitem [{\citenamefont {Tabak}\ \emph {et~al.}(2005)\citenamefont {Tabak},
  \citenamefont {Clark}, \citenamefont {Hatchett}, \citenamefont {Key},
  \citenamefont {Lasinski}, \citenamefont {Snavely}, \citenamefont {Wilks},
  \citenamefont {Town}, \citenamefont {Stephens}, \citenamefont {Campbell},
  \citenamefont {Kodama}, \citenamefont {Mima}, \citenamefont {Tanaka},
  \citenamefont {Atzeni},\ and\ \citenamefont {Freeman}}]{Tabak2005}%
  \BibitemOpen
  \bibfield  {author} {\bibinfo {author} {\bibfnamefont {M.}~\bibnamefont
  {Tabak}}, \bibinfo {author} {\bibfnamefont {D.~S.}\ \bibnamefont {Clark}},
  \bibinfo {author} {\bibfnamefont {S.~P.}\ \bibnamefont {Hatchett}}, \bibinfo
  {author} {\bibfnamefont {M.~H.}\ \bibnamefont {Key}}, \bibinfo {author}
  {\bibfnamefont {B.~F.}\ \bibnamefont {Lasinski}}, \bibinfo {author}
  {\bibfnamefont {R.~A.}\ \bibnamefont {Snavely}}, \bibinfo {author}
  {\bibfnamefont {S.~C.}\ \bibnamefont {Wilks}}, \bibinfo {author}
  {\bibfnamefont {R.~P.~J.}\ \bibnamefont {Town}}, \bibinfo {author}
  {\bibfnamefont {R.}~\bibnamefont {Stephens}}, \bibinfo {author}
  {\bibfnamefont {E.~M.}\ \bibnamefont {Campbell}}, \bibinfo {author}
  {\bibfnamefont {R.}~\bibnamefont {Kodama}}, \bibinfo {author} {\bibfnamefont
  {K.}~\bibnamefont {Mima}}, \bibinfo {author} {\bibfnamefont {K.~A.}\
  \bibnamefont {Tanaka}}, \bibinfo {author} {\bibfnamefont {S.}~\bibnamefont
  {Atzeni}},\ and\ \bibinfo {author} {\bibfnamefont {R.}~\bibnamefont
  {Freeman}},\ }\bibfield  {title} {\bibinfo {title} {Review of progress in
  fast ignition},\ }\href {https://doi.org/10.1063/1.1871246} {\bibfield
  {journal} {\bibinfo  {journal} {Physics of Plasmas}\ }\textbf {\bibinfo
  {volume} {12}},\ \bibinfo {pages} {057305} (\bibinfo {year}
  {2005})}\BibitemShut {NoStop}%
\bibitem [{\citenamefont {Zhang}\ \emph {et~al.}(2020)\citenamefont {Zhang},
  \citenamefont {Wang}, \citenamefont {Yang}, \citenamefont {Wu}, \citenamefont
  {Ma}, \citenamefont {Jiao}, \citenamefont {Zhang}, \citenamefont {Wu},
  \citenamefont {Yuan}, \citenamefont {Li},\ and\ \citenamefont
  {Zhu}}]{Zhang2020}%
  \BibitemOpen
  \bibfield  {author} {\bibinfo {author} {\bibfnamefont {J.}~\bibnamefont
  {Zhang}}, \bibinfo {author} {\bibfnamefont {W.~M.}\ \bibnamefont {Wang}},
  \bibinfo {author} {\bibfnamefont {X.~H.}\ \bibnamefont {Yang}}, \bibinfo
  {author} {\bibfnamefont {D.}~\bibnamefont {Wu}}, \bibinfo {author}
  {\bibfnamefont {Y.~Y.}\ \bibnamefont {Ma}}, \bibinfo {author} {\bibfnamefont
  {J.~L.}\ \bibnamefont {Jiao}}, \bibinfo {author} {\bibfnamefont
  {Z.}~\bibnamefont {Zhang}}, \bibinfo {author} {\bibfnamefont {F.~Y.}\
  \bibnamefont {Wu}}, \bibinfo {author} {\bibfnamefont {X.~H.}\ \bibnamefont
  {Yuan}}, \bibinfo {author} {\bibfnamefont {Y.~T.}\ \bibnamefont {Li}},\ and\
  \bibinfo {author} {\bibfnamefont {J.~Q.}\ \bibnamefont {Zhu}},\ }\bibfield
  {title} {\bibinfo {title} {Double-cone ignition scheme for inertial
  confinement fusion},\ }\href {https://doi.org/10.1098/rsta.2020.0015}
  {\bibfield  {journal} {\bibinfo  {journal} {Philosophical Transactions of the
  Royal Society A: Mathematical, Physical and Engineering Sciences}\ }\textbf
  {\bibinfo {volume} {378}},\ \bibinfo {pages} {20200015} (\bibinfo {year}
  {2020})}\BibitemShut {NoStop}%
\bibitem [{\citenamefont {Zylstra}\ \emph {et~al.}(2022)\citenamefont
  {Zylstra}, \citenamefont {Hurricane}, \citenamefont {Callahan}, \citenamefont
  {Kritcher}, \citenamefont {Ralph}, \citenamefont {Robey}, \citenamefont
  {Ross}, \citenamefont {Young}, \citenamefont {Baker}, \citenamefont {Casey}
  \emph {et~al.}}]{zylstra2022burning}%
  \BibitemOpen
  \bibfield  {author} {\bibinfo {author} {\bibfnamefont {A.}~\bibnamefont
  {Zylstra}}, \bibinfo {author} {\bibfnamefont {O.}~\bibnamefont {Hurricane}},
  \bibinfo {author} {\bibfnamefont {D.}~\bibnamefont {Callahan}}, \bibinfo
  {author} {\bibfnamefont {A.}~\bibnamefont {Kritcher}}, \bibinfo {author}
  {\bibfnamefont {J.}~\bibnamefont {Ralph}}, \bibinfo {author} {\bibfnamefont
  {H.}~\bibnamefont {Robey}}, \bibinfo {author} {\bibfnamefont
  {J.}~\bibnamefont {Ross}}, \bibinfo {author} {\bibfnamefont {C.}~\bibnamefont
  {Young}}, \bibinfo {author} {\bibfnamefont {K.}~\bibnamefont {Baker}},
  \bibinfo {author} {\bibfnamefont {D.}~\bibnamefont {Casey}}, \emph {et~al.},\
  }\bibfield  {title} {\bibinfo {title} {Burning plasma achieved in inertial
  fusion},\ }\href {https://doi.org/https://doi.org/10.1038/s41586-021-04281-w}
  {\bibfield  {journal} {\bibinfo  {journal} {Nature}\ }\textbf {\bibinfo
  {volume} {601}},\ \bibinfo {pages} {542} (\bibinfo {year}
  {2022})}\BibitemShut {NoStop}%
\bibitem [{\citenamefont {Abu-Shawareb}\ \emph {et~al.}(2024)\citenamefont
  {Abu-Shawareb}, \citenamefont {Acree}, \citenamefont {Adams}, \citenamefont
  {Adams}, \citenamefont {Addis}, \citenamefont {Aden} \emph
  {et~al.}}]{Abu2024Achievement}%
  \BibitemOpen
  \bibfield  {author} {\bibinfo {author} {\bibfnamefont {H.}~\bibnamefont
  {Abu-Shawareb}}, \bibinfo {author} {\bibfnamefont {R.}~\bibnamefont {Acree}},
  \bibinfo {author} {\bibfnamefont {P.}~\bibnamefont {Adams}}, \bibinfo
  {author} {\bibfnamefont {J.}~\bibnamefont {Adams}}, \bibinfo {author}
  {\bibfnamefont {B.}~\bibnamefont {Addis}}, \bibinfo {author} {\bibfnamefont
  {R.}~\bibnamefont {Aden}}, \emph {et~al.},\ }\bibfield  {title} {\bibinfo
  {title} {Achievement of target gain larger than unity in an inertial fusion
  experiment},\ }\href {https://doi.org/10.1103/PhysRevLett.132.065102}
  {\bibfield  {journal} {\bibinfo  {journal} {Phys. Rev. Lett.}\ }\textbf
  {\bibinfo {volume} {132}},\ \bibinfo {pages} {065102} (\bibinfo {year}
  {2024})}\BibitemShut {NoStop}%
\bibitem [{\citenamefont {Rinderknecht}\ \emph {et~al.}(2014)\citenamefont
  {Rinderknecht}, \citenamefont {Sio}, \citenamefont {Li}, \citenamefont
  {Zylstra}, \citenamefont {Rosenberg}, \citenamefont {Amendt}, \citenamefont
  {Delettrez}, \citenamefont {Bellei}, \citenamefont {Frenje}, \citenamefont
  {Gatu~Johnson}, \citenamefont {S\'eguin}, \citenamefont {Petrasso},
  \citenamefont {Betti}, \citenamefont {Glebov}, \citenamefont {Meyerhofer},
  \citenamefont {Sangster}, \citenamefont {Stoeckl}, \citenamefont {Landen},
  \citenamefont {Smalyuk}, \citenamefont {Wilks}, \citenamefont {Greenwood},\
  and\ \citenamefont {Nikroo}}]{Rinderknecht2014}%
  \BibitemOpen
  \bibfield  {author} {\bibinfo {author} {\bibfnamefont {H.~G.}\ \bibnamefont
  {Rinderknecht}}, \bibinfo {author} {\bibfnamefont {H.}~\bibnamefont {Sio}},
  \bibinfo {author} {\bibfnamefont {C.~K.}\ \bibnamefont {Li}}, \bibinfo
  {author} {\bibfnamefont {A.~B.}\ \bibnamefont {Zylstra}}, \bibinfo {author}
  {\bibfnamefont {M.~J.}\ \bibnamefont {Rosenberg}}, \bibinfo {author}
  {\bibfnamefont {P.}~\bibnamefont {Amendt}}, \bibinfo {author} {\bibfnamefont
  {J.}~\bibnamefont {Delettrez}}, \bibinfo {author} {\bibfnamefont
  {C.}~\bibnamefont {Bellei}}, \bibinfo {author} {\bibfnamefont {J.~A.}\
  \bibnamefont {Frenje}}, \bibinfo {author} {\bibfnamefont {M.}~\bibnamefont
  {Gatu~Johnson}}, \bibinfo {author} {\bibfnamefont {F.~H.}\ \bibnamefont
  {S\'eguin}}, \bibinfo {author} {\bibfnamefont {R.~D.}\ \bibnamefont
  {Petrasso}}, \bibinfo {author} {\bibfnamefont {R.}~\bibnamefont {Betti}},
  \bibinfo {author} {\bibfnamefont {V.~Y.}\ \bibnamefont {Glebov}}, \bibinfo
  {author} {\bibfnamefont {D.~D.}\ \bibnamefont {Meyerhofer}}, \bibinfo
  {author} {\bibfnamefont {T.~C.}\ \bibnamefont {Sangster}}, \bibinfo {author}
  {\bibfnamefont {C.}~\bibnamefont {Stoeckl}}, \bibinfo {author} {\bibfnamefont
  {O.}~\bibnamefont {Landen}}, \bibinfo {author} {\bibfnamefont {V.~A.}\
  \bibnamefont {Smalyuk}}, \bibinfo {author} {\bibfnamefont {S.}~\bibnamefont
  {Wilks}}, \bibinfo {author} {\bibfnamefont {A.}~\bibnamefont {Greenwood}},\
  and\ \bibinfo {author} {\bibfnamefont {A.}~\bibnamefont {Nikroo}},\
  }\bibfield  {title} {\bibinfo {title} {First observations of nonhydrodynamic
  mix at the fuel-shell interface in shock-driven inertial confinement
  implosions},\ }\href {https://doi.org/10.1103/PhysRevLett.112.135001}
  {\bibfield  {journal} {\bibinfo  {journal} {Phys. Rev. Lett.}\ }\textbf
  {\bibinfo {volume} {112}},\ \bibinfo {pages} {135001} (\bibinfo {year}
  {2014})}\BibitemShut {NoStop}%
\bibitem [{\citenamefont {Rinderknecht}\ \emph {et~al.}(2018)\citenamefont
  {Rinderknecht}, \citenamefont {Amendt}, \citenamefont {Wilks},\ and\
  \citenamefont {Collins}}]{rinderknecht2018kinetic}%
  \BibitemOpen
  \bibfield  {author} {\bibinfo {author} {\bibfnamefont {H.~G.}\ \bibnamefont
  {Rinderknecht}}, \bibinfo {author} {\bibfnamefont {P.~A.}\ \bibnamefont
  {Amendt}}, \bibinfo {author} {\bibfnamefont {S.~C.}\ \bibnamefont {Wilks}},\
  and\ \bibinfo {author} {\bibfnamefont {G.}~\bibnamefont {Collins}},\
  }\bibfield  {title} {\bibinfo {title} {Kinetic physics in icf: present
  understanding and future directions},\ }\href
  {https://doi.org/10.1088/1361-6587/aab79f} {\bibfield  {journal} {\bibinfo
  {journal} {Plasma Physics and Controlled Fusion}\ }\textbf {\bibinfo {volume}
  {60}},\ \bibinfo {pages} {064001} (\bibinfo {year} {2018})}\BibitemShut
  {NoStop}%
\bibitem [{\citenamefont {Yin}\ \emph {et~al.}(2019)\citenamefont {Yin},
  \citenamefont {Albright}, \citenamefont {Vold}, \citenamefont {Nystrom},
  \citenamefont {Bird},\ and\ \citenamefont {Bowers}}]{Yin2019}%
  \BibitemOpen
  \bibfield  {author} {\bibinfo {author} {\bibfnamefont {L.}~\bibnamefont
  {Yin}}, \bibinfo {author} {\bibfnamefont {B.~J.}\ \bibnamefont {Albright}},
  \bibinfo {author} {\bibfnamefont {E.~L.}\ \bibnamefont {Vold}}, \bibinfo
  {author} {\bibfnamefont {W.~D.}\ \bibnamefont {Nystrom}}, \bibinfo {author}
  {\bibfnamefont {R.~F.}\ \bibnamefont {Bird}},\ and\ \bibinfo {author}
  {\bibfnamefont {K.~J.}\ \bibnamefont {Bowers}},\ }\bibfield  {title}
  {\bibinfo {title} {Plasma kinetic effects on interfacial mix and burn rates
  in multispatial dimensions},\ }\href {https://doi.org/10.1063/1.5109257}
  {\bibfield  {journal} {\bibinfo  {journal} {Physics of Plasmas}\ }\textbf
  {\bibinfo {volume} {26}},\ \bibinfo {pages} {062302} (\bibinfo {year}
  {2019})}\BibitemShut {NoStop}%
\bibitem [{\citenamefont {Hartouni}\ \emph {et~al.}(2023)\citenamefont
  {Hartouni}, \citenamefont {Moore}, \citenamefont {Crilly}, \citenamefont
  {Appelbe}, \citenamefont {Amendt}, \citenamefont {Baker}, \citenamefont
  {Casey}, \citenamefont {Clark}, \citenamefont {D{\"o}ppner}, \citenamefont
  {Eckart} \emph {et~al.}}]{hartouni2023evidence}%
  \BibitemOpen
  \bibfield  {author} {\bibinfo {author} {\bibfnamefont {E.}~\bibnamefont
  {Hartouni}}, \bibinfo {author} {\bibfnamefont {A.}~\bibnamefont {Moore}},
  \bibinfo {author} {\bibfnamefont {A.}~\bibnamefont {Crilly}}, \bibinfo
  {author} {\bibfnamefont {B.}~\bibnamefont {Appelbe}}, \bibinfo {author}
  {\bibfnamefont {P.}~\bibnamefont {Amendt}}, \bibinfo {author} {\bibfnamefont
  {K.}~\bibnamefont {Baker}}, \bibinfo {author} {\bibfnamefont
  {D.}~\bibnamefont {Casey}}, \bibinfo {author} {\bibfnamefont
  {D.}~\bibnamefont {Clark}}, \bibinfo {author} {\bibfnamefont
  {T.}~\bibnamefont {D{\"o}ppner}}, \bibinfo {author} {\bibfnamefont
  {M.}~\bibnamefont {Eckart}}, \emph {et~al.},\ }\bibfield  {title} {\bibinfo
  {title} {Evidence for suprathermal ion distribution in burning plasmas},\
  }\href {https://doi.org/10.1038/s41567-022-01809-3} {\bibfield  {journal}
  {\bibinfo  {journal} {Nature physics}\ }\textbf {\bibinfo {volume} {19}},\
  \bibinfo {pages} {72} (\bibinfo {year} {2023})}\BibitemShut {NoStop}%
\bibitem [{\citenamefont {Higginson}\ \emph {et~al.}(2025)\citenamefont
  {Higginson}, \citenamefont {Izumi}, \citenamefont {Rosen}, \citenamefont
  {Volegov}, \citenamefont {Chapman}, \citenamefont {Fittinghoff} \emph
  {et~al.}}]{Higginson2025Evidence}%
  \BibitemOpen
  \bibfield  {author} {\bibinfo {author} {\bibfnamefont {D.~P.}\ \bibnamefont
  {Higginson}}, \bibinfo {author} {\bibfnamefont {N.}~\bibnamefont {Izumi}},
  \bibinfo {author} {\bibfnamefont {M.~D.}\ \bibnamefont {Rosen}}, \bibinfo
  {author} {\bibfnamefont {P.}~\bibnamefont {Volegov}}, \bibinfo {author}
  {\bibfnamefont {T.}~\bibnamefont {Chapman}}, \bibinfo {author} {\bibfnamefont
  {D.~N.}\ \bibnamefont {Fittinghoff}}, \emph {et~al.},\ }\bibfield  {title}
  {\bibinfo {title} {Direct evidence of multispecies hydrodynamics in
  ignition-scale hohlraums},\ }\href
  {https://doi.org/10.1103/PhysRevLett.134.165101} {\bibfield  {journal}
  {\bibinfo  {journal} {Phys. Rev. Lett.}\ }\textbf {\bibinfo {volume} {134}},\
  \bibinfo {pages} {165101} (\bibinfo {year} {2025})}\BibitemShut {NoStop}%
\bibitem [{\citenamefont {Murphy}(2014)}]{Murphy2014}%
  \BibitemOpen
  \bibfield  {author} {\bibinfo {author} {\bibfnamefont {T.~J.}\ \bibnamefont
  {Murphy}},\ }\bibfield  {title} {\bibinfo {title} {The effect of turbulent
  kinetic energy on inferred ion temperature from neutron spectra},\ }\href
  {https://doi.org/10.1063/1.4885342} {\bibfield  {journal} {\bibinfo
  {journal} {Physics of Plasmas}\ }\textbf {\bibinfo {volume} {21}},\ \bibinfo
  {pages} {072701} (\bibinfo {year} {2014})}\BibitemShut {NoStop}%
\bibitem [{\citenamefont {Mannion}\ \emph {et~al.}(2023)\citenamefont
  {Mannion}, \citenamefont {Taitano}, \citenamefont {Appelbe}, \citenamefont
  {Crilly}, \citenamefont {Forrest}, \citenamefont {Glebov}, \citenamefont
  {Knauer}, \citenamefont {McKenty}, \citenamefont {Mohamed}, \citenamefont
  {Stoeckl}, \citenamefont {Keenan}, \citenamefont {Chittenden}, \citenamefont
  {Adrian}, \citenamefont {Frenje}, \citenamefont {Kabadi}, \citenamefont
  {Gatu~Johnson},\ and\ \citenamefont {Regan}}]{Mannion2023}%
  \BibitemOpen
  \bibfield  {author} {\bibinfo {author} {\bibfnamefont {O.~M.}\ \bibnamefont
  {Mannion}}, \bibinfo {author} {\bibfnamefont {W.~T.}\ \bibnamefont
  {Taitano}}, \bibinfo {author} {\bibfnamefont {B.~D.}\ \bibnamefont
  {Appelbe}}, \bibinfo {author} {\bibfnamefont {A.~J.}\ \bibnamefont {Crilly}},
  \bibinfo {author} {\bibfnamefont {C.~J.}\ \bibnamefont {Forrest}}, \bibinfo
  {author} {\bibfnamefont {V.~Y.}\ \bibnamefont {Glebov}}, \bibinfo {author}
  {\bibfnamefont {J.~P.}\ \bibnamefont {Knauer}}, \bibinfo {author}
  {\bibfnamefont {P.~W.}\ \bibnamefont {McKenty}}, \bibinfo {author}
  {\bibfnamefont {Z.~L.}\ \bibnamefont {Mohamed}}, \bibinfo {author}
  {\bibfnamefont {C.}~\bibnamefont {Stoeckl}}, \bibinfo {author} {\bibfnamefont
  {B.~D.}\ \bibnamefont {Keenan}}, \bibinfo {author} {\bibfnamefont {J.~P.}\
  \bibnamefont {Chittenden}}, \bibinfo {author} {\bibfnamefont
  {P.}~\bibnamefont {Adrian}}, \bibinfo {author} {\bibfnamefont
  {J.}~\bibnamefont {Frenje}}, \bibinfo {author} {\bibfnamefont
  {N.}~\bibnamefont {Kabadi}}, \bibinfo {author} {\bibfnamefont
  {M.}~\bibnamefont {Gatu~Johnson}},\ and\ \bibinfo {author} {\bibfnamefont
  {S.~P.}\ \bibnamefont {Regan}},\ }\bibfield  {title} {\bibinfo {title}
  {Evidence of non-maxwellian ion velocity distributions in spherical
  shock-driven implosions},\ }\href
  {https://doi.org/10.1103/PhysRevE.108.035201} {\bibfield  {journal} {\bibinfo
   {journal} {Phys. Rev. E}\ }\textbf {\bibinfo {volume} {108}},\ \bibinfo
  {pages} {035201} (\bibinfo {year} {2023})}\BibitemShut {NoStop}%
\bibitem [{\citenamefont {Taitano}\ \emph {et~al.}(2018)\citenamefont
  {Taitano}, \citenamefont {Simakov}, \citenamefont {Chacón},\ and\
  \citenamefont {Keenan}}]{Taitano2018}%
  \BibitemOpen
  \bibfield  {author} {\bibinfo {author} {\bibfnamefont {W.~T.}\ \bibnamefont
  {Taitano}}, \bibinfo {author} {\bibfnamefont {A.~N.}\ \bibnamefont
  {Simakov}}, \bibinfo {author} {\bibfnamefont {L.}~\bibnamefont {Chacón}},\
  and\ \bibinfo {author} {\bibfnamefont {B.}~\bibnamefont {Keenan}},\
  }\bibfield  {title} {\bibinfo {title} {Yield degradation in
  inertial-confinement-fusion implosions due to shock-driven kinetic
  fuel-species stratification and viscous heating},\ }\href
  {https://doi.org/10.1063/1.5024402} {\bibfield  {journal} {\bibinfo
  {journal} {Physics of Plasmas}\ }\textbf {\bibinfo {volume} {25}},\ \bibinfo
  {pages} {056310} (\bibinfo {year} {2018})}\BibitemShut {NoStop}%
\bibitem [{\citenamefont {Petschek}\ and\ \citenamefont
  {Henderson}(1979)}]{Petschek_1979}%
  \BibitemOpen
  \bibfield  {author} {\bibinfo {author} {\bibfnamefont {A.}~\bibnamefont
  {Petschek}}\ and\ \bibinfo {author} {\bibfnamefont {D.}~\bibnamefont
  {Henderson}},\ }\bibfield  {title} {\bibinfo {title} {Influence of
  high-energy ion loss on dt reaction rate in laser fusion pellets},\ }\href
  {https://doi.org/10.1088/0029-5515/19/12/012} {\bibfield  {journal} {\bibinfo
   {journal} {Nuclear Fusion}\ }\textbf {\bibinfo {volume} {19}},\ \bibinfo
  {pages} {1678} (\bibinfo {year} {1979})}\BibitemShut {NoStop}%
\bibitem [{\citenamefont {Molvig}\ \emph {et~al.}(2012)\citenamefont {Molvig},
  \citenamefont {Hoffman}, \citenamefont {Albright}, \citenamefont {Nelson},\
  and\ \citenamefont {Webster}}]{Molvig2012Knudsen}%
  \BibitemOpen
  \bibfield  {author} {\bibinfo {author} {\bibfnamefont {K.}~\bibnamefont
  {Molvig}}, \bibinfo {author} {\bibfnamefont {N.~M.}\ \bibnamefont {Hoffman}},
  \bibinfo {author} {\bibfnamefont {B.~J.}\ \bibnamefont {Albright}}, \bibinfo
  {author} {\bibfnamefont {E.~M.}\ \bibnamefont {Nelson}},\ and\ \bibinfo
  {author} {\bibfnamefont {R.~B.}\ \bibnamefont {Webster}},\ }\bibfield
  {title} {\bibinfo {title} {Knudsen layer reduction of fusion reactivity},\
  }\href {https://doi.org/10.1103/PhysRevLett.109.095001} {\bibfield  {journal}
  {\bibinfo  {journal} {Phys. Rev. Lett.}\ }\textbf {\bibinfo {volume} {109}},\
  \bibinfo {pages} {095001} (\bibinfo {year} {2012})}\BibitemShut {NoStop}%
\bibitem [{\citenamefont {Albright}\ \emph {et~al.}(2013)\citenamefont
  {Albright}, \citenamefont {Molvig}, \citenamefont {Huang}, \citenamefont
  {Simakov}, \citenamefont {Dodd}, \citenamefont {Hoffman}, \citenamefont
  {Kagan},\ and\ \citenamefont {Schmit}}]{Albright2013}%
  \BibitemOpen
  \bibfield  {author} {\bibinfo {author} {\bibfnamefont {B.~J.}\ \bibnamefont
  {Albright}}, \bibinfo {author} {\bibfnamefont {K.}~\bibnamefont {Molvig}},
  \bibinfo {author} {\bibfnamefont {C.-K.}\ \bibnamefont {Huang}}, \bibinfo
  {author} {\bibfnamefont {A.~N.}\ \bibnamefont {Simakov}}, \bibinfo {author}
  {\bibfnamefont {E.~S.}\ \bibnamefont {Dodd}}, \bibinfo {author}
  {\bibfnamefont {N.~M.}\ \bibnamefont {Hoffman}}, \bibinfo {author}
  {\bibfnamefont {G.}~\bibnamefont {Kagan}},\ and\ \bibinfo {author}
  {\bibfnamefont {P.~F.}\ \bibnamefont {Schmit}},\ }\bibfield  {title}
  {\bibinfo {title} {Revised knudsen-layer reduction of fusion reactivity},\
  }\href {https://doi.org/10.1063/1.4833639} {\bibfield  {journal} {\bibinfo
  {journal} {Physics of Plasmas}\ }\textbf {\bibinfo {volume} {20}},\ \bibinfo
  {pages} {122705} (\bibinfo {year} {2013})}\BibitemShut {NoStop}%
\bibitem [{\citenamefont {Schmit}\ \emph {et~al.}(2013)\citenamefont {Schmit},
  \citenamefont {Molvig},\ and\ \citenamefont {Nakhleh}}]{Schmit2013}%
  \BibitemOpen
  \bibfield  {author} {\bibinfo {author} {\bibfnamefont {P.~F.}\ \bibnamefont
  {Schmit}}, \bibinfo {author} {\bibfnamefont {K.}~\bibnamefont {Molvig}},\
  and\ \bibinfo {author} {\bibfnamefont {C.~W.}\ \bibnamefont {Nakhleh}},\
  }\bibfield  {title} {\bibinfo {title} {Tail-ion transport and knudsen layer
  formation in the presence of magnetic fields},\ }\href
  {https://doi.org/10.1063/1.4831958} {\bibfield  {journal} {\bibinfo
  {journal} {Physics of Plasmas}\ }\textbf {\bibinfo {volume} {20}},\ \bibinfo
  {pages} {112705} (\bibinfo {year} {2013})}\BibitemShut {NoStop}%
\bibitem [{\citenamefont {Fisher}\ \emph {et~al.}(1994)\citenamefont {Fisher},
  \citenamefont {Parks}, \citenamefont {McChesney},\ and\ \citenamefont
  {Rosenbluth}}]{Fisher_1994}%
  \BibitemOpen
  \bibfield  {author} {\bibinfo {author} {\bibfnamefont {R.}~\bibnamefont
  {Fisher}}, \bibinfo {author} {\bibfnamefont {P.}~\bibnamefont {Parks}},
  \bibinfo {author} {\bibfnamefont {J.}~\bibnamefont {McChesney}},\ and\
  \bibinfo {author} {\bibfnamefont {M.}~\bibnamefont {Rosenbluth}},\ }\bibfield
   {title} {\bibinfo {title} {Fast alpha particle diagnostics using knock-on
  ion tails},\ }\href {https://doi.org/10.1088/0029-5515/34/10/I01} {\bibfield
  {journal} {\bibinfo  {journal} {Nuclear Fusion}\ }\textbf {\bibinfo {volume}
  {34}},\ \bibinfo {pages} {1291} (\bibinfo {year} {1994})}\BibitemShut
  {NoStop}%
\bibitem [{\citenamefont {Ballabio}\ \emph {et~al.}(1997)\citenamefont
  {Ballabio}, \citenamefont {Gorini},\ and\ \citenamefont
  {K\"allne}}]{Ballabio1997}%
  \BibitemOpen
  \bibfield  {author} {\bibinfo {author} {\bibfnamefont {L.}~\bibnamefont
  {Ballabio}}, \bibinfo {author} {\bibfnamefont {G.}~\bibnamefont {Gorini}},\
  and\ \bibinfo {author} {\bibfnamefont {J.}~\bibnamefont {K\"allne}},\
  }\bibfield  {title} {\bibinfo {title} {\ensuremath{\alpha}-particle knock-on
  signature in the neutron emission of dt plasmas},\ }\href
  {https://doi.org/10.1103/PhysRevE.55.3358} {\bibfield  {journal} {\bibinfo
  {journal} {Phys. Rev. E}\ }\textbf {\bibinfo {volume} {55}},\ \bibinfo
  {pages} {3358} (\bibinfo {year} {1997})}\BibitemShut {NoStop}%
\bibitem [{\citenamefont {Xue}\ \emph {et~al.}(2025)\citenamefont {Xue},
  \citenamefont {Wu},\ and\ \citenamefont {Zhang}}]{XUE2025359}%
  \BibitemOpen
  \bibfield  {author} {\bibinfo {author} {\bibfnamefont {Y.}~\bibnamefont
  {Xue}}, \bibinfo {author} {\bibfnamefont {D.}~\bibnamefont {Wu}},\ and\
  \bibinfo {author} {\bibfnamefont {J.}~\bibnamefont {Zhang}},\ }\bibfield
  {title} {\bibinfo {title} {Mechanisms behind the surprising observation of
  supra-thermal ions in nif’s fusion burning plasmas},\ }\href
  {https://doi.org/https://doi.org/10.1016/j.scib.2024.11.050} {\bibfield
  {journal} {\bibinfo  {journal} {Science Bulletin}\ }\textbf {\bibinfo
  {volume} {70}},\ \bibinfo {pages} {359} (\bibinfo {year} {2025})}\BibitemShut
  {NoStop}%
\bibitem [{\citenamefont {Fetsch}\ and\ \citenamefont
  {Fisch}(2025{\natexlab{a}})}]{fetsch2025enhancement}%
  \BibitemOpen
  \bibfield  {author} {\bibinfo {author} {\bibfnamefont {H.}~\bibnamefont
  {Fetsch}}\ and\ \bibinfo {author} {\bibfnamefont {N.~J.}\ \bibnamefont
  {Fisch}},\ }\bibfield  {title} {\bibinfo {title} {Enhancement to fusion
  reactivity in sheared flows},\ }\href {https://doi.org/10.1103/5nll-y8rx}
  {\bibfield  {journal} {\bibinfo  {journal} {Phys. Rev. Lett.}\ }\textbf
  {\bibinfo {volume} {135}},\ \bibinfo {pages} {155101} (\bibinfo {year}
  {2025}{\natexlab{a}})}\BibitemShut {NoStop}%
\bibitem [{\citenamefont {Fetsch}\ and\ \citenamefont
  {Fisch}(2025{\natexlab{b}})}]{fetsch2025analytical}%
  \BibitemOpen
  \bibfield  {author} {\bibinfo {author} {\bibfnamefont {H.}~\bibnamefont
  {Fetsch}}\ and\ \bibinfo {author} {\bibfnamefont {N.~J.}\ \bibnamefont
  {Fisch}},\ }\bibfield  {title} {\bibinfo {title} {Analytical models for the
  enhancement of fusion reactivity by turbulence},\ }\href
  {https://doi.org/10.1063/5.0285620} {\bibfield  {journal} {\bibinfo
  {journal} {Physics of Plasmas}\ }\textbf {\bibinfo {volume} {32}},\ \bibinfo
  {pages} {112703} (\bibinfo {year} {2025}{\natexlab{b}})}\BibitemShut
  {NoStop}%
\bibitem [{\citenamefont {Rieger}\ and\ \citenamefont
  {Duffy}(2004)}]{Rieger_2004}%
  \BibitemOpen
  \bibfield  {author} {\bibinfo {author} {\bibfnamefont {F.~M.}\ \bibnamefont
  {Rieger}}\ and\ \bibinfo {author} {\bibfnamefont {P.}~\bibnamefont {Duffy}},\
  }\bibfield  {title} {\bibinfo {title} {Shear acceleration in relativistic
  astrophysical jets},\ }\href {https://doi.org/10.1086/425167} {\bibfield
  {journal} {\bibinfo  {journal} {The Astrophysical Journal}\ }\textbf
  {\bibinfo {volume} {617}},\ \bibinfo {pages} {155} (\bibinfo {year}
  {2004})}\BibitemShut {NoStop}%
\bibitem [{\citenamefont {Bosch}\ and\ \citenamefont
  {Hale}(1992)}]{Bosch_1992}%
  \BibitemOpen
  \bibfield  {author} {\bibinfo {author} {\bibfnamefont {H.-S.}\ \bibnamefont
  {Bosch}}\ and\ \bibinfo {author} {\bibfnamefont {G.}~\bibnamefont {Hale}},\
  }\bibfield  {title} {\bibinfo {title} {Improved formulas for fusion
  cross-sections and thermal reactivities},\ }\href
  {https://doi.org/10.1088/0029-5515/32/4/I07} {\bibfield  {journal} {\bibinfo
  {journal} {Nuclear Fusion}\ }\textbf {\bibinfo {volume} {32}},\ \bibinfo
  {pages} {611} (\bibinfo {year} {1992})}\BibitemShut {NoStop}%
\bibitem [{\citenamefont {Rosenbluth}\ \emph {et~al.}(1957)\citenamefont
  {Rosenbluth}, \citenamefont {MacDonald},\ and\ \citenamefont
  {Judd}}]{Rosenbluth1957}%
  \BibitemOpen
  \bibfield  {author} {\bibinfo {author} {\bibfnamefont {M.~N.}\ \bibnamefont
  {Rosenbluth}}, \bibinfo {author} {\bibfnamefont {W.~M.}\ \bibnamefont
  {MacDonald}},\ and\ \bibinfo {author} {\bibfnamefont {D.~L.}\ \bibnamefont
  {Judd}},\ }\bibfield  {title} {\bibinfo {title} {Fokker-planck equation for
  an inverse-square force},\ }\href {https://doi.org/10.1103/PhysRev.107.1}
  {\bibfield  {journal} {\bibinfo  {journal} {Phys. Rev.}\ }\textbf {\bibinfo
  {volume} {107}},\ \bibinfo {pages} {1} (\bibinfo {year} {1957})}\BibitemShut
  {NoStop}%
\bibitem [{\citenamefont {Bhatnagar}\ \emph {et~al.}(1954)\citenamefont
  {Bhatnagar}, \citenamefont {Gross},\ and\ \citenamefont {Krook}}]{BGK1954}%
  \BibitemOpen
  \bibfield  {author} {\bibinfo {author} {\bibfnamefont {P.~L.}\ \bibnamefont
  {Bhatnagar}}, \bibinfo {author} {\bibfnamefont {E.~P.}\ \bibnamefont
  {Gross}},\ and\ \bibinfo {author} {\bibfnamefont {M.}~\bibnamefont {Krook}},\
  }\bibfield  {title} {\bibinfo {title} {A model for collision processes in
  gases. i. small amplitude processes in charged and neutral one-component
  systems},\ }\href {https://doi.org/10.1103/PhysRev.94.511} {\bibfield
  {journal} {\bibinfo  {journal} {Phys. Rev.}\ }\textbf {\bibinfo {volume}
  {94}},\ \bibinfo {pages} {511} (\bibinfo {year} {1954})}\BibitemShut
  {NoStop}%
\bibitem [{\citenamefont {Press}\ \emph {et~al.}(2007)\citenamefont {Press},
  \citenamefont {Teukolsky}, \citenamefont {Vetterling},\ and\ \citenamefont
  {Flannery}}]{Press2007}%
  \BibitemOpen
  \bibfield  {author} {\bibinfo {author} {\bibfnamefont {W.~H.}\ \bibnamefont
  {Press}}, \bibinfo {author} {\bibfnamefont {S.~A.}\ \bibnamefont
  {Teukolsky}}, \bibinfo {author} {\bibfnamefont {W.~T.}\ \bibnamefont
  {Vetterling}},\ and\ \bibinfo {author} {\bibfnamefont {B.~P.}\ \bibnamefont
  {Flannery}},\ }\href@noop {} {\emph {\bibinfo {title} {Numerical Recipes 3rd
  Edition: The Art of Scientific Computing}}},\ \bibinfo {edition} {3rd}\ ed.\
  (\bibinfo  {publisher} {Cambridge University Press},\ \bibinfo {address}
  {USA},\ \bibinfo {year} {2007})\BibitemShut {NoStop}%
\bibitem [{\citenamefont {Wu}\ and\ \citenamefont {Zhang}(2023)}]{Wu2023}%
  \BibitemOpen
  \bibfield  {author} {\bibinfo {author} {\bibfnamefont {D.}~\bibnamefont
  {Wu}}\ and\ \bibinfo {author} {\bibfnamefont {J.}~\bibnamefont {Zhang}},\
  }\bibfield  {title} {\bibinfo {title} {Head-on collision of large-scale high
  density plasmas jets: A first-principle kinetic simulation approach},\ }\href
  {https://doi.org/10.1063/5.0149413} {\bibfield  {journal} {\bibinfo
  {journal} {Physics of Plasmas}\ }\textbf {\bibinfo {volume} {30}},\ \bibinfo
  {pages} {072711} (\bibinfo {year} {2023})}\BibitemShut {NoStop}%
\bibitem [{\citenamefont {Le}\ \emph {et~al.}(2023)\citenamefont {Le},
  \citenamefont {Stanier}, \citenamefont {Yin}, \citenamefont {Wetherton},
  \citenamefont {Keenan},\ and\ \citenamefont {Albright}}]{10.1063/5.0146529}%
  \BibitemOpen
  \bibfield  {author} {\bibinfo {author} {\bibfnamefont {A.}~\bibnamefont
  {Le}}, \bibinfo {author} {\bibfnamefont {A.}~\bibnamefont {Stanier}},
  \bibinfo {author} {\bibfnamefont {L.}~\bibnamefont {Yin}}, \bibinfo {author}
  {\bibfnamefont {B.}~\bibnamefont {Wetherton}}, \bibinfo {author}
  {\bibfnamefont {B.}~\bibnamefont {Keenan}},\ and\ \bibinfo {author}
  {\bibfnamefont {B.}~\bibnamefont {Albright}},\ }\bibfield  {title} {\bibinfo
  {title} {Hybrid-vpic: An open-source kinetic/fluid hybrid particle-in-cell
  code},\ }\href {https://doi.org/10.1063/5.0146529} {\bibfield  {journal}
  {\bibinfo  {journal} {Physics of Plasmas}\ }\textbf {\bibinfo {volume}
  {30}},\ \bibinfo {pages} {063902} (\bibinfo {year} {2023})}\BibitemShut
  {NoStop}%
\bibitem [{\citenamefont {Takizuka}\ and\ \citenamefont
  {Abe}(1977)}]{TAKIZUKA_1977}%
  \BibitemOpen
  \bibfield  {author} {\bibinfo {author} {\bibfnamefont {T.}~\bibnamefont
  {Takizuka}}\ and\ \bibinfo {author} {\bibfnamefont {H.}~\bibnamefont {Abe}},\
  }\bibfield  {title} {\bibinfo {title} {A binary collision model for plasma
  simulation with a particle code},\ }\href
  {https://doi.org/https://doi.org/10.1016/0021-9991(77)90099-7} {\bibfield
  {journal} {\bibinfo  {journal} {Journal of Computational Physics}\ }\textbf
  {\bibinfo {volume} {25}},\ \bibinfo {pages} {205} (\bibinfo {year}
  {1977})}\BibitemShut {NoStop}%
\bibitem [{\citenamefont {Nanbu}\ and\ \citenamefont
  {Yonemura}(1998)}]{Nanbu_1998}%
  \BibitemOpen
  \bibfield  {author} {\bibinfo {author} {\bibfnamefont {K.}~\bibnamefont
  {Nanbu}}\ and\ \bibinfo {author} {\bibfnamefont {S.}~\bibnamefont
  {Yonemura}},\ }\bibfield  {title} {\bibinfo {title} {Weighted particles in
  coulomb collision simulations based on the theory of a cumulative scattering
  angle},\ }\href {https://doi.org/https://doi.org/10.1006/jcph.1998.6049}
  {\bibfield  {journal} {\bibinfo  {journal} {Journal of Computational
  Physics}\ }\textbf {\bibinfo {volume} {145}},\ \bibinfo {pages} {639}
  (\bibinfo {year} {1998})}\BibitemShut {NoStop}%
\bibitem [{\citenamefont {Pérez}\ \emph {et~al.}(2012)\citenamefont {Pérez},
  \citenamefont {Gremillet}, \citenamefont {Decoster}, \citenamefont {Drouin},\
  and\ \citenamefont {Lefebvre}}]{Perez2012}%
  \BibitemOpen
  \bibfield  {author} {\bibinfo {author} {\bibfnamefont {F.}~\bibnamefont
  {Pérez}}, \bibinfo {author} {\bibfnamefont {L.}~\bibnamefont {Gremillet}},
  \bibinfo {author} {\bibfnamefont {A.}~\bibnamefont {Decoster}}, \bibinfo
  {author} {\bibfnamefont {M.}~\bibnamefont {Drouin}},\ and\ \bibinfo {author}
  {\bibfnamefont {E.}~\bibnamefont {Lefebvre}},\ }\bibfield  {title} {\bibinfo
  {title} {Improved modeling of relativistic collisions and collisional
  ionization in particle-in-cell codes},\ }\href
  {https://doi.org/10.1063/1.4742167} {\bibfield  {journal} {\bibinfo
  {journal} {Physics of Plasmas}\ }\textbf {\bibinfo {volume} {19}},\ \bibinfo
  {pages} {083104} (\bibinfo {year} {2012})}\BibitemShut {NoStop}%
\end{thebibliography}%

\end{document}